\documentclass[a4paper,11pt]{article}
\usepackage{jheppub}
\usepackage{graphicx}
\usepackage{epsfig}
\usepackage{amsmath, amssymb}
\usepackage{dcolumn}
\usepackage{bm}
\usepackage{amsfonts}

\usepackage{subfigure}
\usepackage{amssymb}
\usepackage{epstopdf}

\def\dj{\hbox{d\kern-0.347em \vrule width 0.3em height 1.252ex depth
-1.21ex \kern 0.051em}}

\def\Re{{\rm Re\,}}

\newcommand{\be}{\begin{equation}}
\newcommand{\ee}{\end{equation}}
\newcommand{\ben}{\begin{equation*}}
\newcommand{\een}{\end{equation*}}
\newcommand{\bea}{\begin{eqnarray}}
\newcommand{\eea}{\end{eqnarray}}
\newcommand{\bean}{\begin{eqnarray*}}
\newcommand{\eean}{\end{eqnarray*}}
\newcommand{\brr}{\begin{array}}
\newcommand{\err}{\end{array}}
\newcommand{\bc}{\begin{center}}
\newcommand{\ec}{\end{center}}
\newcommand{\nn}{\nonumber }

\newcommand{\bk}{{\mathbf k}}

\newcommand{\bx}{{\mathbf x}}

\newcommand{\bn}{{\mathbf n}}
\newcommand{\tn}{{\tilde{\bn}}}
\newcommand{\bem}{{\mathbf m}}
\newcommand{\keff}{{\bk_{\rm eff}}}

\newcommand{\TT}{{\rm TT}}

\newcommand{\dx}{\ensuremath{\delta x}}
\newcommand{\hI}{\hspace{1cm}}
\newcommand{\hVII}{\hspace{.7cm}}
\newcommand{\hV}{\hspace{.5cm}}

\title{On the Transverse-Traceless Projection in Lattice Simulations of Gravitational Wave Production}

\author[a]{Daniel G.~Figueroa}
\affiliation[a]{Physics Department, University of Helsinki and Helsinki Institute of Physics\\
P.O. Box 64, FI-00014, Helsinki, Finland}
\emailAdd{daniel.figueroa@helsinki.fi}
\author[b]{Juan Garc\'ia-Bellido}
\affiliation[b]{Instituto de F\'isica Te\'orica UAM/CSIC, Universidad Aut\'onoma de Madrid\\ Cantoblanco 28049 Madrid, Spain}
\emailAdd{juan.garciabellido@uam.es}
\author[c]{Arttu Rajantie}
\affiliation[c]{Theoretical Physics Group, Department of Physics, Imperial College London,\\
London SW7 2AZ, United Kingdom}
\emailAdd{a.rajantie@imperial.ac.uk}

\date{\today}

\abstract{
It has recently been pointed out that the usual procedure employed in order to obtain the transverse-traceless (TT) part of metric perturbations in lattice simulations was inconsistent with the fact that those fields live in the lattice and not in the continuum. It was claimed that this could lead to a larger amplitude and a wrong shape for the gravitational wave (GW) spectra obtained in numerical simulations of (p)reheating. In order to address this issue, we have defined a consistent prescription in the lattice for extracting the TT part of the metric perturbations. We demonstrate explicitly that the GW spectra obtained with the old continuum-based TT projection only differ marginally in amplitude and shape with respect to the new lattice-based ones. We conclude that one can therefore trust the predictions appearing in the literature on the spectra of GW produced during (p)reheating and similar scenarios simulated on a lattice.}

\begin{document}
\maketitle

\section{Introduction}

Gravitational waves (GW) are expected to be produced copiously in the early universe in processes like (p)reheating after inflation~\cite{KTGW,JuanGW,EL,GBF,DBFKU,DufauxHybrid,DFGB}, phase transitions~\cite{PhaseTransitions,Apreda,Nicolis,Caprini}, during the turbulent motion of plasmas~\cite{Turbulence} and from the self-ordering dynamics of Goldstone fields~\cite{FFRGB,JSKM}. As opposed to the GW background generated during inflation from the quantum fluctuations of the metric~\cite{Starobinsky}, these post-inflationary processes correspond to classical mechanisms of GW production, due to the motion of large overdensities.

Such backgrounds of GW could open a new window into the early universe by providing precious information about the dynamics responsible for its production, much before primordial nucleosynthesis. The very violent dynamics of the fields sourcing the GW cannot be described in linear perturbation theory, and usually takes place very far from equilibrium. This is the reason why the study of the GW production in the early universe is usually done with the help of lattice simulations. It is therefore of great importance to have a precise handle on those simulations to be sure that the predictions made on the amplitude and shape of the GW spectra is accurate enough for the future GW observatories~\cite{LIGO} to detect and constrain these backgrounds.

In homogeneous and isotropic background spaces, the six (independent) physical degrees of freedom of the metric split into irreducible representations of rotations. There are two scalar, two vector and two tensor perturbations. The two tensor components correspond to the two polarizations of the GW, the transverse and traceless (TT) degrees of freedom ({\it d.o.f.}) of the metric perturbations. The flat Friedman-Robertson-Walker (FRW) line element with metric perturbations in the synchronous gauge, can be written as
\begin{equation}
ds^2 = - dt^2 + a^2(t)\Big(\delta_{ij} + h_{ij}({\bf x},t)\Big)dx^idx^j\,,
\end{equation} 
The equations of motion of TT {\it d.o.f.} of $h_{ij}$ are
\begin{equation}\label{hijeom}
\partial_\mu\partial^\mu h_{ij}^{\TT} = 16\pi G\,\Pi_{ij}^{\TT}\,,
\end{equation} 
with $\Pi_{ij}^{\TT}$ the transverse-traceless part of the full anisotropic-stress tensor $\Pi_{ij}$.

The transverse-traceless tensor tensor $\Pi_{ij}^{\TT}$ is obtained by applying a projector 
$\Lambda_{ij,lm}$ in momentum space
\begin{equation}\label{eq:ContinuumProjector0}
\tilde\Pi_{ij}^\TT(\bk,t) = \Lambda_{ij,lm}(\hat\bk)\,\tilde\Pi_{lm}(\bk,t)
\end{equation}
where
\begin{eqnarray}\label{eq:Lambda}
\Lambda_{ij,lm}(\hat\bk) \equiv P_{il}(\hat\bk)P_{jm}(\hat\bk) - \frac{1}{2}P_{ij}(\hat\bk)P_{lm}(\hat\bk)\,,\\\label{eq:Pij}
P_{ij}(\hat\bk) \equiv \delta_{ij} - \hat k_i\hat k_j\hI\hI,
\end{eqnarray}
and $\hat\bk=\bk/|\bk|$. 

Because the projector is non-local in space, applying it to the source $\Pi_{ij}$ at every time step is computationally expensive. In practice, it is therefore more convenient and still mathematically equivalent (see~\cite{GBF} for details) to consider a tensor $h_{ij}$ which satisfies the equation of motion
\begin{equation}\label{uijeom}
\partial_\mu\partial^\mu h_{ij} = 16\pi G\,\Pi_{ij}\,,
\end{equation}
and apply the projector
\begin{equation}\label{eq:ContinuumProjector}
{\tilde h}_{ij}^\TT(\bk,t) = \Lambda_{ij,lm}(\hat\bk)\,\tilde h_{lm}(\bk,t)
\end{equation}
only when calculating the output. Fourier transforming back ${\tilde h}_{ij}^\TT(\bk,t)$ to coordinate space, one finds that the metric perturbation 
\begin{equation}
h_{ij}^\TT(\bx,t) = \int \frac{d\bk}{(2\pi)^3} e^{-i\bk\bx}\Lambda_{ij,lm}(\hat\bk)\tilde h_{lm}(\bk,t)
\end{equation} 
verifies the required conditions
\begin{eqnarray}
h_{ji}^{\TT} &=& h_{ij}^{\TT},\quad\mbox{(Symmetry)}\label{equ:symcondition}\\
\sum_i h_{ii}^{\TT}&=&0,\quad\mbox{(Tracelessness)}\label{equ:tracecondition}\\
\sum_i\nabla_i h_{ij}^{\TT}&=&0,\quad\mbox{(Transversality)}\label{equ:transcondition}
\end{eqnarray}
at all $\bx, t$, necessary for identifying $h_{ij}^\TT(\bx,t)$ with the gravitational wave {\it d.o.f.} 
It is also easy to check that the projector (\ref{eq:Lambda}) is maximal in the sense that it leaves any tensor $a^\TT_{ij}$ that satisfies the conditions (\ref{equ:symcondition})--(\ref{equ:transcondition}) unchanged,
\begin{equation}
\Lambda_{ij,lm}a^\TT_{lm}=a^\TT_{ij}. \quad\mbox{(Maximality)}\label{equ:maxcondition}
\end{equation}
This guarantees that we capture the whole transverse-traceless component.

The above procedure to obtain the TT {\it d.o.f.} is well-defined in the continuum. 
However, it has recently been pointed out in Ref.~\cite{Huang2011} that on a lattice one needs to pay more attention to the definition of the projector~(\ref{eq:Lambda}), since one can define the momentum $\bk$ in many different ways. In particular, Ref.~\cite{Huang2011} claims that if one applies the wrong projector, a significant leak of scalar modes into the tensor modes (GW) might occur, significantly modifying the amplitude of the GW spectrum. In the context of (p)reheating, all lattice simulations carried out in recent years by the different groups~\cite{EL,GBF,DBFKU,DufauxHybrid,DFGB}, see also Refs.~\cite{Caltech,BasteroGil,Jedamzik:2010hq}, filtered the $\TT$ metric {\it d.o.f.} with the projector defined above. Therefore, whether such projector is or not appropriate for the lattice, challenges the validity of the GW spectra shown in the literature. 

In this paper we investigate this issue in detail. We will review first, in Section~\ref{sec:Discretization}, some ideas about the discretization aspects in a lattice, and then in Section~\ref{sec:TTlatticeProjection} we will present our method for obtaining a TT-projector consistent with the symmetries of the lattice. In Section~\ref{sec:TTlatticeProjectionResults} we will compare analytically and numerically several GW spectra obtained with the continuum-based projection and with different lattice-based projections. Finally in Section~\ref{sec:discussion} we will summarize and conclude.

\section{Lattice discretization}
\label{sec:Discretization}

\subsection{The lattice derivative}

When one simulates the dynamics of non-equilibrium fields like in (p)reheating, the field equations are discretized on a lattice. 
We consider a lattice with $N^3$ points (representing spatial comoving coordinates) labeled as ${\bf n} = (n_1,n_2,n_3)$, with $n_i = 0,1,...,N-1$. A function $f(\bx)$ in space is represented by a lattice function $f({\bf n})$ which has the same value as $f(\bx)$ at $\bx=\bn\, \dx$, with $\dx \equiv L/N$ the lattice spacing, and $L$ the length of the lattice. 

To discretize the equations of motion, one has to replace the continuum derivative with a lattice expression that has the same continuum limit. A simple and symmetric definition of a lattice derivative is the neutral one,
\begin{equation}
\label{equ:neutrald}
[\nabla^0_i f]({\bf n}) = \frac{f({\bf n}+\hat\imath) - f({\bf n}-\hat\imath)}{2\dx},
\end{equation}
where $\hat\imath$ is the unit vector in direction $i$.
This has the drawback that it is insensitive to variations at the smallest scale of one lattice spacing.
In this sense, a definition involving the nearest neighbors may be preferable.
A common way to do this, is to define
the forward and backward derivatives
\begin{equation}
\label{equ:forwardbackwardd}
[\nabla^\pm_i f]({\bf n}) = \frac{\pm f({\bf n}\pm \hat\imath) \mp f({\bf n})}{\dx},
\end{equation}
but these definitions lack the symmetry of Eq.~(\ref{equ:neutrald}).
This issue can be solved by defining the derivative at half-way between the lattice sites, as
\begin{equation}
\label{equ:nnd}
[\nabla_i f](\mathbf{n}+\hat{\imath}/2)=\frac{f(\mathbf{n}+\hat{\imath})-f(\mathbf{n})}{\delta x}.
\end{equation}
To improve accuracy, one can also consider lattice derivatives which involve more points, but in practice the definitions have a symmetry either around a lattice site as Eq.~(\ref{equ:neutrald}) or half-way between lattice sites as~(\ref{equ:nnd}).

In order to extract the transverse-traceless component of $h_{ij}$, one needs to apply the lattice version of the projector (\ref{eq:Lambda}). Since the projector is defined in Fourier space, on the lattice one has to use the discrete Fourier transform (DFT) $\tilde f(\tilde{\bf n})$, defined as
\begin{eqnarray}
f({\bf n}) = \frac{1}{N^3}\sum_{\tilde n}e^{-\frac{2\pi i}{N}\tilde{\bf n} {\bf n}}\,\tilde f(\tilde{\bf n})\,,\hI \,\tilde f(\tilde{\bf n}) = \sum_{n}e^{+\frac{2\pi i}{N}\tilde{\bf n} {\bf n}} f({\bf n})\,,
\end{eqnarray}
where the index $\tilde{\bf n} = (\tilde n_1, \tilde n_2, \tilde n_3)$ labels the reciprocal lattice, with $\tilde n_i = -\frac{N}{2}+1, -\frac{N}{2}+2,...$ $-1,0,1,..., \frac{N}{2}$. Imposing periodic boundary conditions in coordinate space, i.e. $f({\bf n} + \hat i N) = f({\bf n})$, there will be necessarily a minimum infrared (IR) momentum $k_{\rm IR} = \frac{2\pi}{L}$ in the Fourier space, such that $\tilde{\bf n}$ will be representing the continuum momentum $\bk = (\tilde n_1, \tilde n_2, \tilde n_3)\, k_{\rm IR}$. Consequently, there will also be a maximum ultraviolet (UV) momentum $k_{\rm UV} = {N\over2}k_{\rm IR}$ per dimension, and periodic boundary conditions like $\tilde f(\tilde{\bf n} + \hat i N) = \tilde f(\tilde{\bf n})$.

In the calculation of GW production, there are four separate places where one needs to take care of the discretization details of derivatives: in the equations of the fields sourcing the GW, in the two sides of Eq.~(\ref{uijeom}), and in the transverse-traceless projection (\ref{eq:ContinuumProjector}) itself. Ideally one should use a consistent choice of a lattice derivative everywhere, but in some cases there are restrictions that make this very difficult, for example when dealing with gauge fields~\cite{DFGB}. It is enough in any case to have a consistent choice among Eqs.~(\ref{uijeom}) and the equations for the GW sources, whilst the discretization details in the transverse-traceless projection can be considered separately. In this paper we focus precisely on such details of the transverse-traceless projection in a lattice.

\subsection{The transverse-traceless component of the metric}

With the transverse-traceless projection, we want to obtain a tensor $h_{ij}^{\TT}$ that satisfies the three conditions (\ref{equ:symcondition})--(\ref{equ:transcondition}) on the lattice,
with respect to the appropriate lattice derivative. In the literature\footnote{Some papers, for instance Ref.~\cite{EL} and Ref.~\cite{DBFKU}, considered the projection at the level of the source, whereas others like Ref.~\cite{GBF},~\cite{DufauxHybrid} and~\cite{DFGB}, considered the projection at the level of the metric perturbations, as we are discussing here.}, this projection has been done by taking $\Lambda_{ij,lm}(\tilde{\bf n})$ as in Eq.~(\ref{eq:Lambda}) evaluated at $\bk = \tilde{\bf n}\,k_{\rm IR}$,
\begin{equation}\label{eq:ContinuumBasedProjector}
h_{ij}^{\TT}({\bf n}) = \frac{1}{N^3}\sum_{\tilde n}e^{-\frac{2\pi i}{N}\tilde{\bf n} {\bf n}}\Lambda_{ij,lm}(\tilde{\bf n})\,\tilde h_{lm}(\tilde{\bf n}).
\end{equation}
However, as highlighted in Ref.~\cite{Huang2011}, the resulting quantity is not transverse
with respect to any of the usual lattice derivative operators $\nabla_i$, i.e.~$\nabla_ih_{ij}^{\TT} \neq 0$. 

In particular, due to this lack of transversality, Ref.~\cite{Huang2011} claims that a significant leak of scalar modes into the tensor modes might occur in such a way that the amplitude of the GW spectrum extracted with the above continuum-based projector~(\ref{eq:ContinuumBasedProjector}) could be several orders of magnitude higher than it should be.

We therefore need a general and consistent procedure in order to define a TT-projection in the lattice, i.e.~a projector $\Lambda_{ij,lm}$ that restores the transversality with respect to a given lattice derivative. Only then we will be able to quantify the potential distortion of the GW spectra with respect to the results obtained with the continuum-based projector. In order to construct the lattice equivalent of Eq.~(\ref{eq:Lambda}), we need a lattice momentum ${\bf k}$. Such momentum will depend, of course, upon the choice of the lattice derivative with respect to which the transversality condition will be attained. 

\subsection{The lattice momentum}
\label{subsec:LatticeMomenta}

The lattice momentum is given by the Fourier transform of the lattice derivative $\nabla_i$. To keep the discussion general, we do not assume for the moment a specific form for the derivative, but simply assume that it is given by a linear operator in the space of lattice functions. Therefore, the value of the derivative $[\nabla_i f]$ is a linear combination of the field values at different lattice sites,
\begin{equation}
\left[\nabla_i f\right](\bn) = \sum_\bem D_i(\bn,\bem)f(\bem),
\end{equation}
where $D_i(\bn,\bem)$ is a real-valued function of two variables on the lattice.
For example, for the neutral derivative (\ref{equ:neutrald}), we have
\begin{equation}
D^0_i(\bn,\bem)=\frac{\delta_{\bem,\bn+\hat\imath}-\delta_{\bem,\bn-\hat\imath}}{2\dx}.
\end{equation}
Because we want the derivative to be translation invariant, $D_i(\bn,\bem)$ is only a function of the difference $\bn-\bem$, i.e.~$D_i(\bn,\bem)=D_i(\bn-\bem)$, and we can write
\begin{equation}
\left[\nabla_i f\right](\bn) = \sum_\bem D_i(\bn-\bem)f(\bem)
 = \sum_{\bem'} D_i(\bem')f(\bn-\bem') \,.
\end{equation}
For the neutral derivative (\ref{equ:neutrald}), we have
\begin{equation}
D^0_i(\bem')=\frac{\delta_{\bem',-\hat\imath}-\delta_{\bem',\hat\imath}}{2\dx}.
\end{equation}
For the nearest-neighbor derivative (\ref{equ:nnd}), $\bem'$ is half-integer, and one finds
\begin{equation}
D_i(\bem')=\frac{\delta_{\bem',-\hat\imath/2}-\delta_{\bem',\hat\imath/2}}{\dx}.
\end{equation}
More generally, any odd function with compact support will give a meaningful definition of a lattice derivative.

The Fourier transform of the derivative $\nabla_if$  is 
\begin{eqnarray}
\widetilde{\nabla_i\,f}(\tn)&=&\sum_\bn e^{\frac{2\pi i}{N}\tn\cdot\bn}[\nabla_if](\bn)
=\sum_\bn e^{\frac{2\pi i}{N}\tn\cdot\bn}\sum_\bem D_i(\bn-\bem)f(\bem)
\nonumber\\
&=& \sum_{\bn'} e^{\frac{2\pi i}{N}\tn\cdot\bn'}D_i(\bn')\sum_\bem e^{\frac{2\pi i}{N}\tn\cdot\bem}f(\bem)
\equiv -i\keff(\tn)\tilde f(\tn)\,,
\end{eqnarray}
where the effective momentum $\keff(\tn)$ is given by
\begin{equation}
\keff(\tn)=i\sum_\bn e^{\frac{2\pi i}{N}\tn\cdot\bn}D_i(\bn).
\end{equation}
Conversely, any function $\keff(\tn)$ that has the correct leading behaviour in the Taylor expansion of the IR limit $|\tn| \ll N$, i.e.~$\keff(\tn) \approx \tn\,k_{\rm IR}$, defines a lattice derivative through the inverse Fourier transform.

For example, the neutral derivative (\ref{equ:neutrald}) gives
\be\label{eq:neutralMomentum}
k_{{\rm eff},i}^0=\frac{\sin(2\pi \tilde{n}_i/N)}{\dx}\,.
\ee
The forward/backward derivatives (\ref{equ:forwardbackwardd}) give
\be\label{eq:chargedMomentum}
k_{{\rm eff},i}^\pm=2e^{\pm i\pi\tilde{n}_i/N}\frac{\sin(\pi \tilde{n}_i/N)}{\dx} = \frac{\sin(2\pi \tilde{n}_i/N)}{\dx} \pm i
\frac{1-\cos(2\pi \tilde{n}_i/N)}{\dx}\,,
\ee
and the symmetric nearest-neighbor derivative (\ref{equ:nnd}) gives
\be
\label{equ:knnd}
k_{{\rm eff},i}=2\frac{\sin(\pi \tilde{n}_i/N)}{\dx}.
\ee
In general, if the lattice derivative is anti-symmetric, i.e.~$D_{i}(-\bn)=-D_i(\bn)$, then the lattice momentum $\keff$ is real. This is the case also for the  derivative used in Ref.~\cite{Huang2011}.

\section{The transverse-traceless (TT) projection on the lattice}
\label{sec:TTlatticeProjection}

In this Section we will define 
a lattice projection operator that satisfies the condition (\ref{equ:symcondition})--(\ref{equ:maxcondition}), and which therefore gives
the TT {\it d.o.f.} of metric perturbations living on a lattice. Since the transversality notion on a lattice is associated to the choice of a lattice derivative $\nabla_i$, we will introduce a projector that will guarantee tracelessness and transversality with respect to any $\nabla_i$ chosen.

\subsection{A real TT-projector}
\label{subsec:RealProjector}

Let us start with the simpler case of a real momentum, for example
${\bf k}_{\rm eff}^0$ in Eq.~(\ref{eq:neutralMomentum}). In this case, the projector can be defined in the same way as in continuum.

In analogy with Eqs.~(\ref{eq:Lambda}) and (\ref{eq:Pij}), we define
\begin{equation}
P^0_{ij}(\tilde{\bf n}) = \delta_{ij} - \frac{k^0_{{\rm eff},i}k^0_{{\rm eff},j}}{(k^0_{\rm eff})^2}\,,
\end{equation}
and
\begin{equation}\label{eq:TTprojectorNeutralDerivatives}
 \Lambda_{ij,lm}^0(\tilde\bn) \equiv P^0_{il}(\tilde{\bf n})P^0_{jm}(\tilde{\bf n})-\frac{1}{2}P^0_{ij}(\tilde{\bf n})P^0_{lm}(\tilde{\bf n})\,.
\end{equation}
Using the properties
\begin{equation}
k^0_{{\rm eff},i}P^0_{ij}(\tilde{\bf n}) = 0\,, \hspace{7mm}
P^0_{ij}(\tilde{\bf n})P^0_{jl}(\tilde{\bf n})=P^0_{il}(\tilde{\bf n})\,,\hVII P^0_{ij}(\bn) = P^0_{ji}(\bn),
\end{equation}
it is then straightforward to prove that $\tilde h_{ij}^{\TT}(\tilde{\bf n}) = \Lambda_{ij,lm}^0(\tilde\bn)\tilde h_{lm}(\tilde{\bf n})$ satisfies the required conditions (\ref{equ:symcondition})--(\ref{equ:maxcondition}):\\

\noindent Symmetry:\\
\begin{equation}
\label{equ:symreal}
\Lambda^0_{ji,lm}(\tilde\bn)=\Lambda^0_{ij,ml}(\tilde\bn)
\quad\Rightarrow\quad
\tilde{h}^\TT_{ji}(\tilde\bn) = \tilde{h}^\TT_{ij}(\tilde\bn).
\quad\Rightarrow\quad
h^\TT_{ji}(\bn) = h^\TT_{ij}(\bn).
\end{equation}\\

\noindent Tracelessness: \\
\begin{equation}
\label{equ:tracereal}
\Lambda_{ii,lm}^0(\tilde\bn) = 0 \hV\Rightarrow\hV \tilde h^\TT_{ii}(\tilde{\bf n}) = 0 \hV\Rightarrow\hV h^\TT_{ii}({\bf n}) = 0,~\forall\,\bn
\end{equation}\\

\noindent Transversality:\\
\begin{equation}
\label{equ:transreal}
k^0_{{\rm eff},i}\Lambda_{ij,lm}^0(\tilde\bn) = 0\hV \Rightarrow\hV k^0_{{\rm eff},i}\tilde h^\TT_{ij}(\tilde{\bf n}) = 0 \hV\Rightarrow\hV \nabla^0_i h^\TT_{ij}({\bf n}) = 0,~\forall\,\bn
\end{equation}\\

It is also easy to see that the resulting tensor $h^\TT_{ij}$ is real in coordinate space.
If $h_{ij}(\bn)$ is real, its Fourier transform satisfies $h_{ij}^*(\tilde\bn)=h_{ij}(-\tilde\bn)$, and then because $\Lambda^0_{ij,lm}(\tilde\bn)$ is real and even,
\begin{equation}
h_{ij}^{\TT*}(\tilde\bn)=\Lambda^0_{ij,lm}(\tilde\bn)h^*_{ij}(\tilde\bn)
=\Lambda^{0}_{ij,lm}(-\tilde\bn)h_{ij}(-\tilde\bn)=h_{ij}^{\TT}(-\tilde\bn)
\quad\Rightarrow\quad h^{\TT*}_{ij}(\bn)=h^\TT_{ij}(\bn).
\end{equation}
\\
Finally, to prove the maximality (\ref{equ:maxcondition}) of the operator, we assume a tensor $a^\TT_{ij}$ that satisfies the conditions (\ref{equ:symreal})--(\ref{equ:transreal}), and note that then it also satisfies $P^0_{ij}a^\TT_{jk}=\delta_{ij}a^TT_{jk}$. Therefore we have
\begin{eqnarray}
\Lambda^0_{ij,lm}a^\TT_{lm}=\left(\delta_{il}\delta_{jm}-\frac{1}{2}P^0_{ij}\delta_{lm}\right)
a^\TT_{lm}=a^\TT_{ij}-\frac{1}{2}P^0_{ij}a^\TT_{ll}=a^\TT_{ij},
\end{eqnarray}

Of course, all properties just discussed apply, not only to ${\bf k}^0_{\rm eff}$ in Eq.~(\ref{eq:neutralMomentum}), but to any lattice momentum ${\bf k}_{\rm eff}$ as long it is real. The case of a derivative with an associated real lattice momentum, is therefore a simple generalization of the continuum case.

\subsection{A complex TT-projector}\label{subsec:ComplexProjector}

In the more general case, the lattice momentum is complex. For example, this is the case with the forward/backward derivatives (\ref{equ:forwardbackwardd}) and the associated momenta
${\bf k}^\pm_{\rm eff}$. Thus we will be forced to take a projector $P_{ij}$ that is also complex. 

Thus, we look for a projector $P_{ij}$ that satisfies $k_{{\rm eff},i}P_{ij}({\bf k}) = 0$. In order to do this, we define
\begin{equation}\label{eq:ComplexTransverseProj}
P_{ij}(\tilde{\bf n}) = \delta_{ij} - \frac{(k_{{\rm eff},i})^*k_{{\rm eff},j}}{|\bk_{\rm eff}|^2}\,,
\end{equation}
with $|\bk_{\rm eff}|^2 = {k_{{\rm eff},i}}^*k_{{\rm eff},i}$. This projector is complex and satisfies
\begin{eqnarray}
\begin{tabular}{lccl}
1) $k_{{\rm eff},i}P_{ij}({\bf k}) = 0$ & & & 2) ${k_{{\rm eff},i}^*}P_{ij}({\bf k}) \neq 0$\\
3) $k_{{\rm eff},j}P_{ij}({\bf k}) \neq 0$ & & & 4) ${k_{{\rm eff},j}^*}P_{ij}({\bf k}) = 0$ \\
4) $P^*_{ij}(\tilde{\bf n}) = P_{ji}(\tilde\bn)$ & & & 6) $P_{ij}(-\tilde{\bf n}) = P_{ji}(\tilde\bn)$\\
7) $P_{ij}(\tilde\bn)P_{jl}(\tilde\bn)=P_{il}({\tn})$ & & & 8) $P_{ij}(\tilde\bn)P_{lj}(\tilde\bn) \neq P_{il}({\tn})$
\end{tabular}
\end{eqnarray}
In words, this projector is Hermitian, symmetric under Parity transformations $\bn \leftrightarrow -\bn$, transverse to $\bk$ but not to $\bk^*$, and idempotent ($P^2 = P$) but with no inverse ($\nexists\,P^{-1}$) and non-idempotent modulus ($PP^* \neq P$). Demanding property $1)$ we arrived at the form~(\ref{eq:ComplexTransverseProj}), and then properties $2)-8)$ simply followed from such form. 

If $\Lambda_{ij,lm}(\tilde\bn)$ was built as in the real case (\ref{eq:TTprojectorNeutralDerivatives}), then property $3)$ would prevent $h_{ij}^{\TT}(\tilde\bn) \equiv \Lambda_{ij,lm}(\tilde\bn)h_{lm}(\tilde\bn)$ from being traceless. We are thus forced to redefine also $\Lambda_{ij,lm}$ in order to guarantee the desired TT properties. Moreover, since $P_{ij}$ is now complex, so is $\Lambda_{ij,lm}$. Therefore we must also ensure that $h_{ij}^{\TT}(\bn) = DFT\lbrace\Lambda_{ij,lm}(\tilde\bn)h_{lm}(\tilde\bn)\rbrace$ is real. From the properties of the Fourier transform and demanding $h_{ij}^{\TT^{*}}(\bn) = h_{ij}^{\TT}(\bn)$, the latter condition can be achieved if and only if $\Lambda_{ij,lm}(\tilde\bn)$ satisfies
\begin{equation}
 \Lambda_{ij,lm}^*(\tilde\bn) = \Lambda_{ij,lm}(-\tilde\bn)
\end{equation}
This condition suggests how to build the new projector. We can define
\begin{equation}\label{eq:TTprojectorGeneralDerivatives}
 \Lambda_{ij,lm}(\tilde\bn) = P_{il}(\tilde\bn)P^*_{jm}(\tilde\bn) - \frac{1}{2}P_{ij}(\tilde\bn)P^*_{lm}(\tilde\bn)\,,
\end{equation}
which verifies
\begin{eqnarray}
 \Lambda_{ij,lm}^{*}(\tilde\bn) = \Lambda_{ij,lm}(-\tilde\bn) = \Lambda_{ji,ml}(\tilde\bn) = \Lambda^{*}_{ml,ji}(\tilde\bn) = \Lambda_{lm,ij}(\tilde\bn)\,,
\end{eqnarray} 
From here it is easy to prove that $h_{ij}^{\TT}({\bf n}) \equiv DFT\lbrace\Lambda^+_{ij,lm}(\tilde\bn)h_{lm}(\tilde\bn)\rbrace$ is traceless and transverse, as well as real. For completeness, let us show explicitly how we obtain these conditions 
using~(\ref{eq:TTprojectorGeneralDerivatives}) and its properties:\\

Tracelessness (\ref{equ:tracecondition}):
\begin{eqnarray}
 h_{ii}^{\TT}(\tilde\bn) &=& P_{il}(\tilde\bn)P_{im}^*(\tilde\bn)h_{lm}(\tilde\bn) - \frac{1}{2}P_{ii}(\tilde\bn)P_{lm}^*(\tilde\bn)h_{lm}(\tilde\bn) \nn\\
 &=& P_{ml}(\tilde\bn)h_{lm}(\tilde\bn) - P_{lm}^*(\tilde\bn)h_{lm}(\tilde\bn) = 0
\end{eqnarray} 

Transversality (\ref{equ:transcondition}):
\begin{eqnarray}
 k_{{\rm eff},i}h_{ij}^{\TT}(\tilde\bn) &=& k_{{\rm eff},i}P_{il}(\tilde\bn)P_{jm}^*(\tilde\bn)h_{lm}(\tilde\bn) - \frac{1}{2}k_{{\rm eff},i}P_{ij}(\tilde\bn)P_{lm}^*(\tilde\bn)h_{lm}(\tilde\bn) \nn\\
 &=& 0 - 0 = 0
\end{eqnarray} 

Reality:
\begin{eqnarray}
 && h_{ij}^{\TT^*}(\bn) = \sum_{\tilde\bn} e^{+ik_{\rm IR}\dx\bn\tilde\bn}\Lambda_{ij,lm}^*( \tilde\bn)h_{lm}^*(\tilde\bn) = \sum_{\tilde\bn} e^{+ik_{\rm IR}\dx\bn\tilde\bn}\Lambda_{ij,lm}^*( \tilde\bn)h_{lm}(-\tilde\bn)\nn\\
 && = \sum_{\tilde\bn} e^{-ik_{\rm IR}\dx\bn\tilde\bn}\Lambda_{ij,lm}^*( -\tilde\bn)h_{lm}(\tilde\bn) = \sum_{\tilde\bn} e^{-ik_{\rm IR}\dx\bn\tilde\bn}\Lambda_{ij,lm}( \tilde\bn)h_{lm}(\tilde\bn) \equiv h_{ij}^{\TT}(\bn)\nn\\
\end{eqnarray}

However, we find that the resulting tensor $h^\TT_{ij}$ is not symmetric
, i.e.~$h_{ij}^\TT(\bn) \neq h_{ji}^\TT(\bn)$. Similarly, the following properties, which distinguish between the first and the second index of~$h_{ij}^\TT(n)$, are also verified
\begin{eqnarray}
\begin{tabular}{lccl}
1) $k_{{\rm eff},i}h_{ij}^\TT(\bn) = 0\,,$ & & & 2) ${k_{{\rm eff},i}^*}h_{ij}^\TT(\bn) \neq 0\,,$\\
3) $k_{{\rm eff},j}h_{ij}^\TT(\bn) \neq 0\,,$ & & & 4) ${k_{{\rm eff},j}^*}h_{ij}^\TT(\bn) = 0\,.$ \\
\end{tabular}
\end{eqnarray}
All these asymmetry aspects are simply a consequence of the properties 1)-4) of $P_{ij}$, listed above, which reflects the fact that $P_{ij}$ is not symmetric but rather Hermitian.

Related to these issues we also encounter a subtle aspect about the maximality condition~(\ref{equ:maxcondition}). We find that $\Lambda_{ij,lm}(\bn)A_{lm}^{\TT}(\bn) = A_{ij}^{\TT}(\bn)$ only holds for those transverse-traceless symmetric rank-2 tensors $A_{ij}^{\TT}$ which are transverse, not only with respect the lattice derivative $\nabla_i$ (with lattice momenta $k_{\rm eff}$ used to build $\Lambda_{ij,lm}$), but also with respect the conjugate derivative $\nabla^*_i$ defined through the lattice momentum $k^*_{\rm eff}$. For example, if one builds the projector~(\ref{eq:TTprojectorGeneralDerivatives}) with the lattice momenta associated to forward derivatives $\nabla_i^+$, then $\Lambda_{ij,lm}$ is only maximal with respect those tensors which are transverse both with respect to forward and backward derivatives, i.e.~$\nabla_i^+A_{ij}  = \nabla_i^-A_{ij} = 0$.

In the IR limit $|\tilde\bn| \ll N$, both $k_{{\rm eff},i}$ and $k_{{\rm eff},i}^*$ approach the same momentum, $\tilde\bn\,k_{\rm IR}$, and thus the full maximality condition and the symmetry under the exchange $i \leftrightarrow j$, are recovered. Thus for arbitrarily big lattices these caveats should not be relevant. In reality, we are of course limited by computer memory and the lattice sizes we can typically consider have no more than $N = 128$, $256$ or $512$ points per dimension, depending upon the field content. Nevertheless, despite these two caveats about the maximality and the even-symmetry, the projector defined by eqs.~(\ref{eq:ComplexTransverseProj}), (\ref{eq:TTprojectorGeneralDerivatives}), is one which generically guarantees reality, transversality and tracelessness on a lattice, and recovers maximality and even-symmetry in the IR limit. Thus, any GW spectra obtained by this method should be reliable at least in the IR region of the Fourier space. 

\subsection{General projector}

To understand the difficulties faced in the complex case, 
let us note that the conditions of symmetry (\ref{equ:symcondition}), tracelessness (\ref{equ:tracecondition}) and transversality (\ref{equ:transcondition}) on a lattice involve comparing and adding together different components of the tensor $h_{ij}^{\TT}$ and its derivatives. For this to be meaningful, these components should be arranged in a symmetric way on the lattice. This is not an issue for derivatives defined on lattice sites, such as the neutral derivative of Eq.~(\ref{equ:neutrald}), which is why the TT projector defined with real lattice momenta verifies nicely all of the required conditions. For those derivatives defined halfway between two lattice sites, with a complex lattice momentum, it is however something to take care of. To illustrate this, let us consider the forward derivative $\nabla_i^+$. The transversality condition becomes
\begin{equation}
\sum_i \left[h_{ij}^{\TT}(\bn+\hat\imath)-h_{ij}^{\TT}(\bn)\right]=0,
\end{equation}
which is not symmetric under parity, because it involves neighboring points only in the positive directions.

Thus, instead of separate asymmetric $\nabla_i$ and $\nabla_i^*$ derivatives with complex lattice momenta, such as the forward and backward derivatives~(\ref{equ:forwardbackwardd}), we should use the symmetric version (\ref{equ:nnd}).
In order to make the lattice transversality condition symmetric under parity, we should then define the tensor $h^\TT_{ij}$ not on the lattice site $\bn$, but at the point $\bn+\hat\imath/2+\hat\jmath/2$, which corresponds to the center of a plaquette spanned by unit vectors $\hat\imath$ and $\hat\jmath$ starting at point $\bn$.
More precisely, off-diagonal components are defined at plaquettes, but the diagonal components, for which $i=j$, live on lattice sites. 
To avoid confusion, we denote the tensor defined in this way by ${\tt h}^\TT_{ij}(\bn+\hat\imath/2+\hat\jmath/2)$.

With this definition,  the symmetry of the tensor, 
\bea\label{eq:sym}
{\tt h}_{ji}^{\TT}(\bn+\hat\imath/2+\hat\jmath/2) = {\tt h}_{ij}^{\TT}(\bn+\hat\imath/2+\hat\jmath/2),
\eea
is a meaningful concept
because the two sides of the equation are defined on the same plaquette $\bn+\hat\imath/2+\hat\jmath/2$.
The same is true for the trace ${\tt h}_{ii}(\mathbf{n})$ because all the terms in the sum are now defined at the same location, 
\bea\label{eq:trace}
\sum_i {\tt h}_{ii}^{\TT}(\bn) = 0,
\eea
and similarly 
in the transversality condition for the tensor ${\tt h}_{ij}$,
\begin{equation}\label{eq:transverse}
[\nabla_i {\tt h}_{ij}](\mathbf{n}+\hat{\jmath}/2)
=\sum_i\frac{{\tt h}_{ij}(\mathbf{n}+\hat{\imath}/2+\hat{\jmath}/2)-{\tt h}_{ij}(\mathbf{n}-\hat{\imath}/2 +\hat{\jmath}/2)}{\delta x}=0,
\end{equation}
all the terms are defined at the same location $\bn+\hat\jmath/2$.

In Fourier space, the transversality condition is just $k_i(\mathbf{\tilde n})\tilde{\tt h}_{ij}(\mathbf{\tilde n})=0$, where $\mathbf{k}(\mathbf{\tilde n})$ is the real momentum in Eq.~(\ref{equ:knnd}) and 
\begin{equation}
\tilde{\tt h}_{ij}(\mathbf{\tilde n})=\sum_\mathbf{n}e^{\frac{2\pi i}{N}\mathbf{\tilde n}\cdot(\mathbf{n}+\hat{\imath}/2+\hat{\jmath}/2)}{\tt h}_{ij}(\mathbf{n}+\hat{\imath}/2+\hat{\jmath}/2).
\end{equation}

Because the momentum $\mathbf{k}(\bn)$ is real, we can now build the transverse-traceless projection in the standard way (\ref{eq:Lambda})--(\ref{eq:Pij}), with
\begin{eqnarray}
\label{equ:realproj}
{\tt \Lambda}_{ij,lm}(\mathbf{\tilde n})={\tt P}_{il}(\mathbf{\tilde n}){\tt P}_{jm}(\mathbf{\tilde n})-\frac{1}{2}{\tt P}_{ij}(\mathbf{\tilde n}){\tt P}_{lm}(\mathbf{\tilde n}),\\
\label{equ:realTransProj}
{\tt P}_{ij}(\mathbf{\tilde n})=\delta_{ij}-\frac{k_ik_j}{k^2}.\hI\hI
\end{eqnarray}
The projected field
\begin{equation}
\tilde{\tt h}_{ij}^{\TT}(\mathbf{\tilde n})={\tt \Lambda}_{ij,lm}(\mathbf{\tilde n})\tilde{\tt h}_{lm}(\mathbf{\tilde n})
\end{equation}
then satisfies obviously all the requirements:
\begin{enumerate}
\item Symmetry: ${\tt \Lambda}_{ji,lm}{\tt h}_{lm}={\tt \Lambda}_{ij,lm}{\tt h}_{lm}$ if ${\tt h}_{ml}={\tt h}_{lm}$. 
\item Reality: ${\tt \Lambda}_{ij,lm}^*(\mathbf{\tilde n})={\tt \Lambda}_{ij,lm}(-\mathbf{\tilde n})$.
\item Tracelessness: ${\tt \Lambda}_{ii,lm}=0$.
\item Transversality: $k_i{\tt \Lambda}_{ij,lm}=0$.
\item Maximality: ${\tt \Lambda}_{ij,lm}{\tt h}_{lm}={\tt h}_{ij}$ for any symmetric, transverse and traceless ${\tt h}_{ij}$.
\end{enumerate}
since the definition of~(\ref{equ:realproj}) is the same as in Section~\ref{subsec:RealProjector}. 

In actual lattice simulations, it is easier to use a tensor $h_{ij}^\TT$ defined on lattice sites. However, now that we have obtained the projector, we can shift the field to lattice sites by a simple translation
\begin{equation}
{h}_{ij}^{\TT}(\bn) \equiv {\tt h}_{ij}^{\TT}(\bn+\hat\imath/2+\hat\jmath/2)\,,
\end{equation}
and derive the form of the equivalent projector in that formulation.
The Fourier transforms of $h^\TT_{ij}$ and ${\tt h}^\TT_{ij}$ are related by
\begin{eqnarray}
\tilde{h}_{ij}^\TT(\mathbf{\tilde n})
&=& \sum_\mathbf{n}e^{\frac{2\pi i}{N}\mathbf{\tilde n}\cdot\mathbf{n}}h_{ij}^\TT(\mathbf{n})
=\sum_\mathbf{n}e^{\frac{2\pi i}{N}\mathbf{\tilde n}\cdot\mathbf{n}}{\tt h}^\TT_{ij}(\mathbf{n}+\hat{\imath}/2+\hat{\jmath}/2) \nn\\
&=& e^{-\frac{\pi i}{N}(\tilde n_i+\tilde n_j)}\tilde{\tt h}^\TT_{ij}(\mathbf{\tilde n}).
\end{eqnarray}
This implies that 
\begin{equation}
\tilde{h}_{ij}^\TT(\mathbf{\tilde n}) = e^{-\frac{\pi i}{N}(\tilde n_i+\tilde n_j)}{\tt \Lambda}_{ij,lm}(\mathbf{\tilde n})e^{+\frac{\pi i}{N}(\tilde n_l+\tilde n_m)}\tilde{h}_{lm}(\mathbf{\tilde n}),
\end{equation}
and therefore, the equivalent projector $\Lambda_{ij,lm}$ for $h^\TT_{ij}$ defined on lattice sites is related to ${\tt \Lambda}_{ij,lm}$, by
\begin{equation}
\Lambda_{ij,lm}(\mathbf{\tilde n}) \equiv e^{-\frac{\pi i}{N}(\tilde n_i+\tilde n_j)}{\tt \Lambda}_{ij,lm}(\mathbf{\tilde n})e^{+\frac{\pi i}{N}(\tilde n_l+\tilde n_m)}
\end{equation}
Noting the relation between $P_{lm}$ defined as a function of the complex lattice momentum\footnote{Here we refer to the complex momentum $k_{{\rm eff},i}$ of the asymmetric derivative $\nabla_i$, which differs from the real momentum $p_{{\rm eff},i}$ of the symmetrized version of $\nabla_i$, just by a complex phase $\varphi_i$, i.e.~$k_{{\rm eff},i}/p_{{\rm eff},i} = e^{+i\varphi_i}$. For the forward derivative this is just $e^{+i\varphi_i} = e^{+i\frac{\pi}{N}\tilde n_i}$.} $\bk_{\rm eff}$, see Section~\ref{subsec:ComplexProjector}, with ${\tt P}_{ij}$ defined previously as a function of the real momentum characteristic of the symmetrized derivative,
\begin{equation}
  P_{ij}(\tilde\bn) = e^{-i\frac{\pi}{N}\tilde n_i}{\tt P}_{ij}(\tilde\bn)e^{+i\frac{\pi}{N}\tilde n_j}\,,
\end{equation}  
then we can write the new projector as 
\begin{eqnarray}
\label{equ:LambdaFinal}
\Lambda_{ij,lm}(\mathbf{\tilde n})
=P_{il}(\mathbf{\tilde n})P_{jm}(\mathbf{\tilde n})-\frac{1}{2}e^{\frac{2\pi i}{N}(\tilde{n}_l-\tilde{n}_j)}P_{ij}(\mathbf{\tilde n})P_{lm}(\mathbf{\tilde n}),\\
\label{eq:ProjTransFinal}
 P_{ij} \equiv \delta_{ij} - \frac{k_{{\rm eff},i}^*k_{{\rm eff},j}}{|k_{{\rm eff}}|^2}\hI\hI\hI
\end{eqnarray}
At the same time, the shift of coordinates also turns Eqs.~(\ref{eq:sym})-(\ref{eq:transverse}) into a less symmetric, but equivalent set of conditions
\begin{eqnarray}
{\rm Tracelessness:}&&\hV\sum_i {h}_{ii}^{\TT}(\bn-\hat\imath) = 0,\nonumber\\
{\rm Transversality:}&&\hV\sum_i\left[{h}_{ij}^{\TT}(\bn)-{h}_{ij}^{\TT}(\bn-\hat\imath)\right] = 0,\nonumber\\
{\rm Symmetry:}&&\hV {h}_{ji}^{\TT}(\bn) = {h}_{ij}^{\TT}(\bn)
\label{equ:latticeconditions2}
\end{eqnarray}

Note that Eq.~(\ref{equ:LambdaFinal}), together with Eq.~(\ref{eq:ProjTransFinal}), define a projector that guarantees maximality and reality of the Fourier transform of the projected {\it d.o.f.} $\tilde h_{ij}(\tilde\bn) = \Lambda_{ij,lm}(\tilde\bn)\tilde h_{lm}(\tilde\bn)$, as well as transversality, tracelessness and even-symmetry, in the way stated in Eqs.~(\ref{equ:latticeconditions2}).
To be more precise, Eqs.~(\ref{equ:LambdaFinal}), (\ref{eq:ProjTransFinal}) will guarantee all these conditions with respect to the forward/backward derivatives defined in~(\ref{equ:forwardbackwardd}), which is the one that we will implement in lattice simulations. For other asymmetric derivatives with lattice momenta $k_{{\rm eff},j}(\tilde\bn) = e^{+i\varphi_j}p_{{\rm eff},j}(\tilde\bn)\,,~\varphi_j \in [0,2\pi),~ p_{{\rm eff},j}(\tilde\bn) \in \Re^+$, but with $e^{+i\varphi_j} \neq e^{+{i\pi\over N}\tilde{n}_j}$, the projector reads
\begin{equation}\label{eq:SuperGeneralProjector}
 \Lambda_{ij,lm}(\tilde\bn) = e^{+i(\Phi_i-\Phi_l)}e^{+i(\Phi_j-\Phi_m)}\left(P_{il}(\mathbf{\tilde n})P_{jm}(\mathbf{\tilde n})-\frac{1}{2}e^{-2(\varphi_j-\varphi_l)}P_{ij}(\mathbf{\tilde n})P_{lm}(\mathbf{\tilde n})\right)\,,
\end{equation}
with $\Phi_i \equiv \varphi_i - {\pi\over N}\tilde n_i$ and $P_{ij}(\tilde\bn)$ defined as in Eq.~(\ref{eq:ProjTransFinal}). Of course, in the case of the forward derivative, $e^{+i\varphi_j} = e^{+i{\pi\over N}\tilde{n}_j}$, so $\Phi_i = 0~\forall~i$, and then eq.~(\ref{eq:SuperGeneralProjector}) reduces to eq.~(\ref{equ:LambdaFinal}). Eq.~(\ref{eq:SuperGeneralProjector}) will guarantee in general the reality and maximality conditions, the even-symmetry and tracelessness defined as in eqs.~(\ref{equ:latticeconditions2}), and transversality with respect to the lattice derivative defined through the momentum $k_{{\rm eff},j}(\tilde\bn) = e^{+i\varphi_j}p_{{\rm eff},j}(\tilde\bn)$.

\section{Comparison of the GW spectra obtained with different projections}
\label{sec:TTlatticeProjectionResults}

In this Section we will show the time evolution of the the GW spectra during (p)reheating in different inflationary models. For each model, we will superimpose the spectra obtained by taking $\Lambda_{ij,lm}(\tilde{\bf n})$ in Eq.~(\ref{eq:Lambda}) evaluated at $\bk = \tilde{\bf n}\,k_{\rm IR}$ (i.e. the continuum-based projector), together with the spectra obtained by using the projectors defined in: 1) Eq.~(\ref{eq:TTprojectorNeutralDerivatives}) evaluated at the momenta of neutral derivatives, 2) Eq.~(\ref{eq:TTprojectorGeneralDerivatives}) evaluated at the lattice momenta of forward derivatives, and 3) Eq.~(\ref{equ:LambdaFinal}) evaluated also at the lattice momenta of forward derivatives. Thus, we will show explicitly how the several procedures defined to extract the GW spectra, compare with each other at different times. We will repeat this exercise for different chaotic and hybrid models of inflation. Before describing the results, however, we must define the GW spectrum. 

The energy density of a homogeneous and isotropic GW background is described in the continuum by~\cite{Maggiore}
\begin{eqnarray}\label{eq:GWrhoCont}
 \rho_{GW}(t) = \int\frac{d\rho_{GW}}{d\log k}\,d\log k &=& \frac{1}{32\pi G}\langle h^{\TT'}_{ij}(\bx,t)h^{\TT'}_{ij}(\bx,t)\rangle\nn\\
 &=& \frac{1}{32\pi G}\frac{1}{V}\int_V d\bx\, \dot h_{ij}(\bx,t)\dot h_{ij}(\bx,t)  \nn\\
&=& \frac{1}{32\pi G}\int\,\,\frac{d\bk d\bk'}{(2\pi)^6} \dot h_{ij}(\bk,t)\dot h_{ij}(\bk',t)\,\frac{1}{V}\int_V d\bx\,e^{-i\bx(\bk-\bk')}\nn\\
\end{eqnarray} 
with $\langle...\rangle$ a spatial average over a sufficiently large volume $V$ encompassing all the relevant wavelengths of the background. In the limit $V^{1/3} \rightarrow \infty$, the spectrum in the continuum (per logarithmic interval) of GW is obtained as
\begin{eqnarray}\label{eq:GWrhoContSpectrum}
\frac{d\rho_{GW}}{d\log k} = 
\frac{k^3}{(4\pi)^3 G\,V} \int \frac{d\Omega_k}{4\pi}\,\dot h_{ij}(k,\hat\bk,t)\dot h_{ij}^*(k,\hat\bk,t)
\end{eqnarray}
where $d\Omega_k$ represent a solid angle element in $\bk$-space. 

In the lattice, we must be more careful since clearly we cannot consider the infinite volume limit. Assuming the volume of the lattice [$V = (Ndx)^3$] encompasses sufficiently well the characteristic wavelengths of the simulated GW background, then it follows
\begin{eqnarray}
 \rho_{GW}(t) &\equiv& \frac{1}{32\pi G}\frac{1}{N^3} \sum_\bn \dot h_{ij}^{\TT}(\bn,t)\dot h_{ij}^{\TT}(\bn,t) \nn\\
 & = & \frac{1}{32\pi G}\frac{1}{N^9} \sum_{\bn}\sum_{\tilde\bn}\sum_{\tilde\bn'} e^{+ik_{IR}dx\bn(\tilde\bn-\tilde\bn')}\dot h_{ij}^{\TT}(\tilde\bn,t)\dot h_{ij}^{\TT^*}(\tilde\bn',t)\nn\\
 & = & \frac{1}{32\pi G}\frac{1}{N^6} \sum_{\tilde\bn} \dot h_{ij}^{\TT}(\tilde\bn,t)\dot h_{ij}^{\TT^*}(\tilde\bn,t)\,,
\end{eqnarray}
where we have used $\sum_\bn e^{+ik_{\rm IR}dx(\tilde\bn-\tilde\bn')\bn} = N^3\,\delta(\tilde\bn-\tilde\bn')$. Binning the momentum-lattice in spherical layers of radii $|\tilde\bn|$ and width $\Delta \tilde n$, $R(\tilde\bn) \equiv \lbrace \tilde\bn' \,/\,\, |\tilde\bn| \leq |\tilde\bn'| < |\tilde\bn|+\Delta\tilde n \rbrace$, then
\begin{eqnarray}\label{eq:GWrhoDiscrete}
\rho_{GW}(t) &=& \frac{1}{32\pi G}\frac{1}{N^6} \sum_{|\tilde\bn|}\sum_{\tilde\bn' \in R({\tilde\bn})}
\dot h_{ij}^{\TT}(\tilde\bn',t)\dot h_{ij}^{\TT^*}(\tilde\bn',t)\nn\\
&=& \frac{1}{32\pi G}\frac{1}{N^6} \sum_{|\tilde\bn|} 4\pi|\tilde\bn|^2
\left\langle\dot h_{ij}^{\TT}(|\tilde\bn|,t)\dot h_{ij}^{\TT^*}(|\tilde\bn|,t)\right\rangle_{R(\tilde\bn)}\nn\\
&=& \sum_{|\tilde\bn|} \left\lbrace\frac{dx^6}{(4\pi)^3 G\,L^3}\,\, k^3(|\tilde\bn|)\,\,
\left\langle\dot h_{ij}^{\TT}(|\tilde\bn|,t)\dot h_{ij}^{\TT^*}(|\tilde\bn|,t)\right\rangle_{R(\tilde\bn)}\right\rbrace\Delta\log k
\end{eqnarray}
where $k(\tilde\bn) \equiv |\tilde\bn|\,$k$_{\rm IR}$, $\Delta\log k \equiv \frac{{\rm k}_{\rm IR}}{k(\tilde\bn)}$, $k_{\rm IR} \equiv 2\pi/L$, $L = N\,dx$, and $\left\langle\dot h_{ij}^{\TT}(|\tilde\bn|,t)\dot h_{ij}^{\TT^*}(|\tilde\bn|,t)\right\rangle_{R(\tilde\bn)}$ is an average over configurations with lattice momenta $\tilde\bn' \in [\,|\tilde\bn|,|\tilde\bn|+\delta\tilde n\,)$.\\

The spectrum of GW in a lattice of volume $V = L^3$, is therefore defined as
\begin{equation}\label{eq:GWrhoDiscreteSpectrum}
\left(\frac{d\rho_{GW}}{d\log k}\right)(\tilde\bn) ~\equiv~ \frac{k^3(|\tilde\bn|)}{(4\pi)^3 G\,L^3}\,\,
\left\langle\left[dx^3\dot h_{ij}^{\TT}(|\tilde\bn|,t)\right]\left[dx^3\dot h_{ij}^{\TT}(|\tilde\bn|,t)\right]^*\right\rangle_{R(\tilde\bn)}
\end{equation} 
In the continuum limit, one identifies\footnote{One is forced to make this identification since the Parseval theorem in the continuum, i.e. the fact that $(2\pi)^3\int d^3\bx\, f^2(\bx)$ = $\int d^3\bk\, |\tilde f(\bk)|^2$, gets translated into $(2\pi)^3dx^3\sum_\bn f^2(\bx)$ = $dk^3\sum_{\tilde\bn} |dx^3 \tilde f(\bk)|^2$ in the lattice, with $dx$ the lattice spacing and $dk = k_{\rm IR}$. In the text we refer to $DFT$ and $CFT$ as the discrete and continuous Fourier transforms respectively.}
$\,DFT\lbrace f(\bn)dx^3 \rbrace \rightarrow CFT\lbrace f(\bx) \rbrace$, and thus, expression~(\ref{eq:GWrhoDiscreteSpectrum}) matches the continuum expression~(\ref{eq:GWrhoContSpectrum}), as it should be. Expression~(\ref{eq:GWrhoDiscreteSpectrum}) highlights that the natural momenta in terms of which to express the lattice GW spectrum, is the continuum one $\bk = \tn k_{\rm IR}$, and not any of the lattice-ones defined from the choice of a lattice derivative.

Having defined the appropriate discretized spectrum of GW, we will now show numerical results from lattice simulations of several scenarios. The key question here will be to show the difference in the GW spectra when considering the different projectors defined in the previous Section. In particular, we will consider chaotic and hybrid models of inflation, since the details of the GW production during reheating in those scenarios have been studied extensively  in the recent years. 

\subsection{Transversality check in lattice simulations}

\begin{figure}
\epsfig{width=5.5cm,height=7.5cm,angle=-90,file=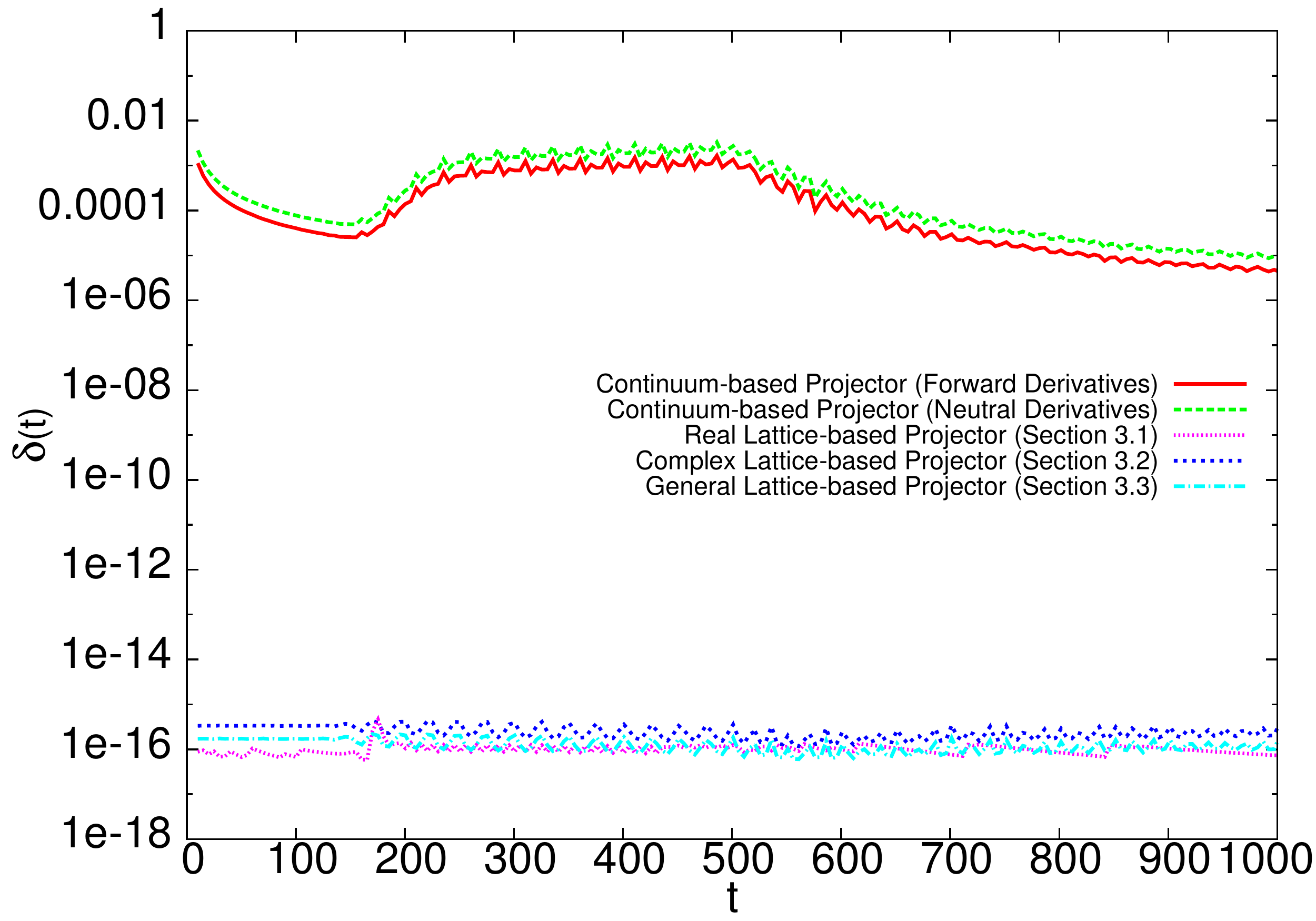}
\epsfig{width=5.5cm,height=7.5cm,angle=-90,file=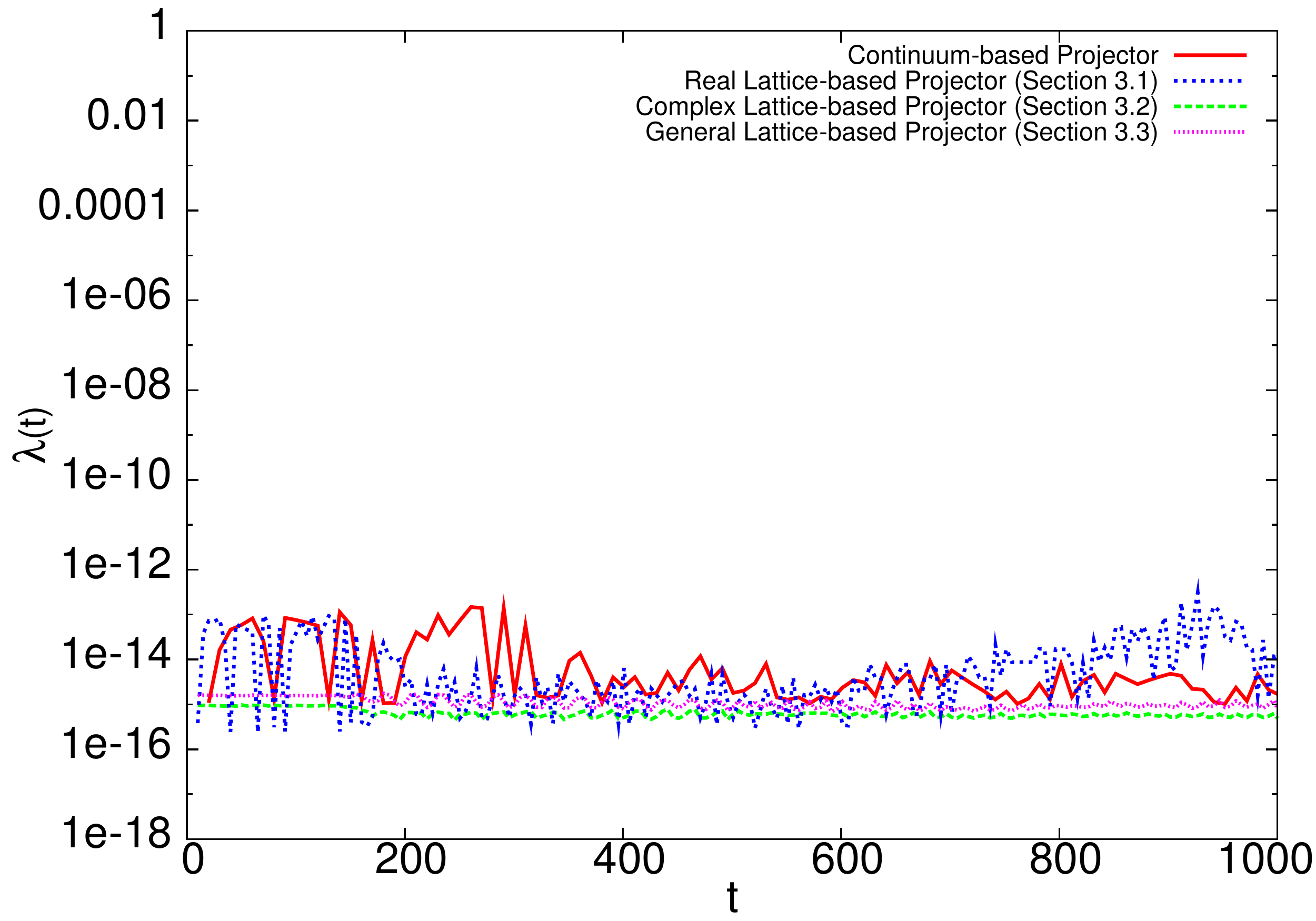}
\caption{{\it Left:} The time evolution of the degree of transversality, $\delta(t)$, obtained for the neutral and forward derivatives, both for the continuum- and lattice-based projectors.  In the latter case, the outcome is clearly limited only by the round-off machine errors, $\delta(t) \sim \mathcal{O}(10^{-16})$, while in the former case, $\delta(t)$ for the continuum can be more than 10 orders of magnitude larger. {\it Right:} The time evolution of $\lambda(t)$ for the same projectors used in the left figure. Here the degree of transversality is well achieved for all cases, including the continuum-based projector. Note that these plots were obtained for a chaotic model, $\frac{\lambda}{4}\phi^4 + \frac{1}{2}g^2\phi^2\chi^2$, with $\lambda/g^2 = 120$. In other models of (p)reheating the curves look very similar, with amplitudes of the same order of magnitude.	
}
\label{fig:deltaANDlambda}
\end{figure}

First we define ${\cal D}_i h^{\rm \TT}_{ij}$ as the sum of $h^\TT_{ij}$ over the lattice sites involved in a particular choice of a derivative scheme $\nabla h_{ij}$. For example, 
\begin{eqnarray}
&& {\cal D}_i h^\TT_{ij}(\bn) \equiv \sum_i(h^\TT_{ij}(\bn) + h^\TT_{ij}(\bn+\hat i))/\delta x,\hV{\rm for~forward~derivatives}\\
&& {\cal D}_i h^\TT_{ij}(\bn) \equiv \sum_i (h^\TT_{ij}(\bn+\hat i) + h^\TT_{ij}(\bn-\hat i))/(2\delta x),\hV{\rm for~neutral~derivatives}
\end{eqnarray}
In the left panel of Fig.~\ref{fig:deltaANDlambda} we show the evolution in time of the dimensionless ratio 
\begin{equation}
\delta(t) \equiv \frac{\langle |\nabla_ih_{ij}^\TT|\rangle}{\langle |{\cal D}_i h^\TT_{ij}| \rangle}
\end{equation} 
for different derivative schemes, where $|\cdots|$ is the absolute value and $\langle\cdots\rangle$ represents an average over all the lattice points. Obviously the value chosen for $j$ here is irrelevant. The evolution of $\delta(t)$ gives an idea of how well the transversality condition $\nabla_i h_{ij}^\TT$ is preserved in the lattice. It is clearly appreciated that with the old-continuum projector $\Lambda_{ij,lm}$, the transversality is not that well achieved. Using the projectors which guarantee transversality with respect to the neutral and forward derivatives, Eqs.~(\ref{eq:TTprojectorNeutralDerivatives}), (\ref{eq:TTprojectorGeneralDerivatives}) and~(\ref{equ:LambdaFinal}), we see that transversality is very well preserved (as it should be, by construction), since $\delta(t)$ is of order $\mathcal{O}(10^{-16})$. Such amplitude simply represents the unavoidable round-off errors of the machine-precision. Thus, it is very clear that the new lattice-based projectors give raise to a well defined notion of transversality for $h_{ij}^\TT$ in the lattice.

We also define the quantity
\begin{equation}
 \lambda(t) \equiv {\langle|\sum_i h_{ii}^\TT|\rangle \over \langle\sum_i|h_{ii}^\TT|\rangle}
\end{equation}
and plot it in the right panel of Figure~\ref{fig:deltaANDlambda}, as obtained for all the same projectors used in the left panel of the same Figure. As expected, for all cases the degree of tracelessness is also very high, only limited again by round-off machine errors. In summary, left and right panels of Fig.~\ref{fig:deltaANDlambda} demonstrate explicitly, and very clearly, that all TT-projectors defined in Section~\ref{sec:TTlatticeProjection}, effectively filter correctly in the lattice the transverse and traceless {\it d.o.f.} of two-rank symmetric tensors.

\subsection{GW spectra in the lattice}

\begin{figure}
\epsfig{width=5.5cm,height=7.5cm,angle=-90,file=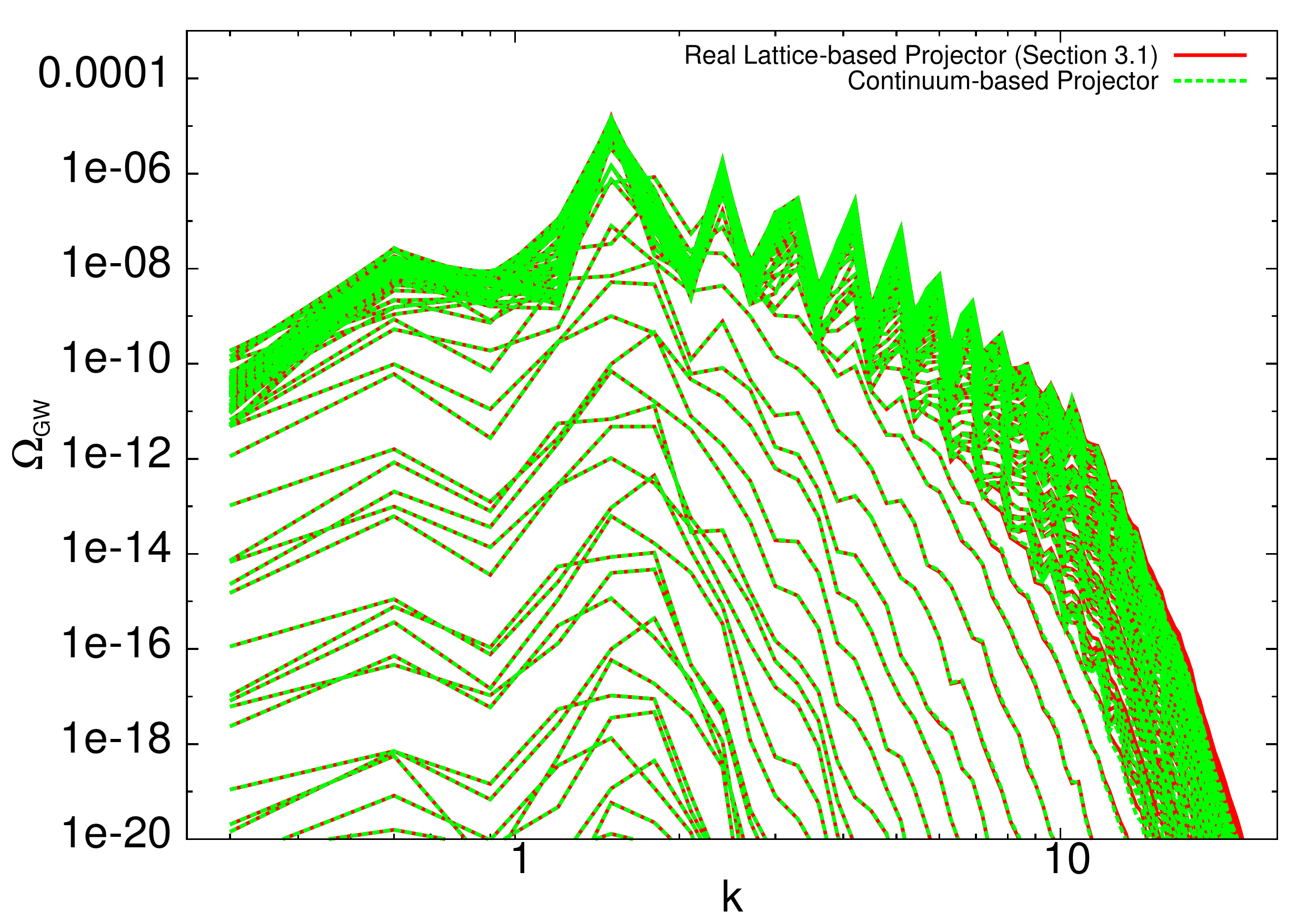}
\epsfig{width=5.5cm,height=7.5cm,angle=-90,file=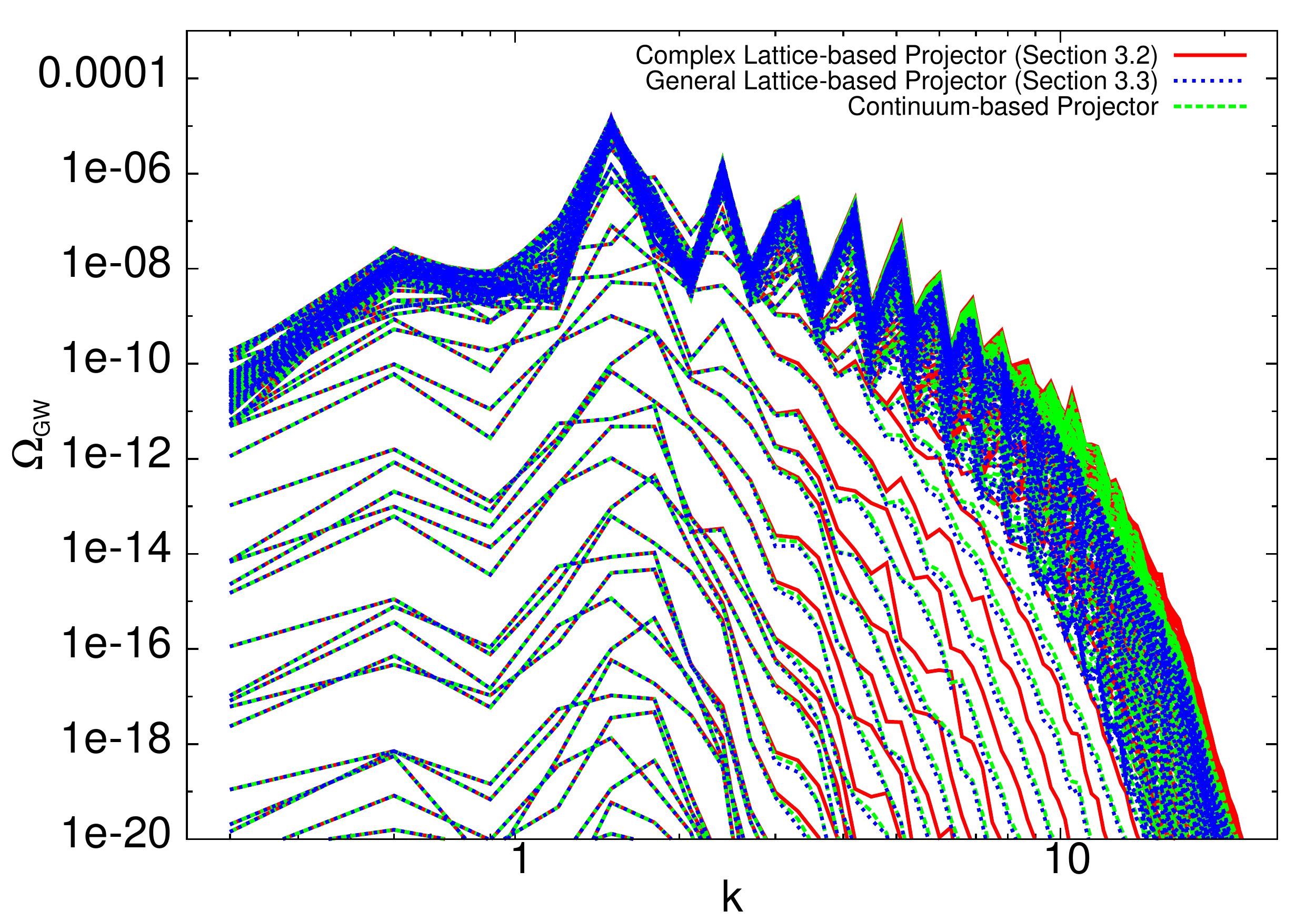}
\caption{The amplitude of the evolution of the GW spectra as obtained with the different projectors defined in Section~\ref{sec:TTlatticeProjection}, during (p)reheating in the model $\frac{\lambda}{4}\phi^4$, with $\lambda = 10^{-13}$ and no coupling to other fields. {\it Left:} Here we compare the spectra obtained with the continuum-based projector versus the one obtained with the real lattice-based projector defined in Eq.~(\ref{eq:TTprojectorNeutralDerivatives}). One can appreciate that the IR part of both spectra are identical during the whole evolution, whereas the UV region shows some difference in amplitude. Such difference smooths out at the end, when the different spectra saturate to their final amplitude. Once the GW production becomes inefficient, and the spectra of GW does not grow further, the difference among the UV tails between the two spectra are simply a factor $\mathcal{O}(1)$. {\it Right}: Here we compare the spectra obtained with the continuum-based projector versus the ones obtained with the complex lattice-based projectors defined in Eq.~(\ref{eq:TTprojectorGeneralDerivatives}) and~Eq.(\ref{equ:LambdaFinal}). Again one appreciates that the IR part of both spectra are identical during the whole evolution. The difference in the UV region when the spectra saturate to their final amplitude, however, is just a factor $\mathcal{O}(10)$, at least at some of the peaks of such UV tail. Note that such discrepancy only appears in a region in $k$-space where the amplitude of the GW spectrum is already a factor $\mathcal{O}(10^{-5})$ smaller than the maximum amplitude. In any case, the difference between the two complex lattice-based projectors is only a factor of order one.}
\label{fig:GW_differenceChaoPure}
\end{figure}

\begin{figure}
\epsfig{width=5.5cm,height=7.5cm,angle=-90,file=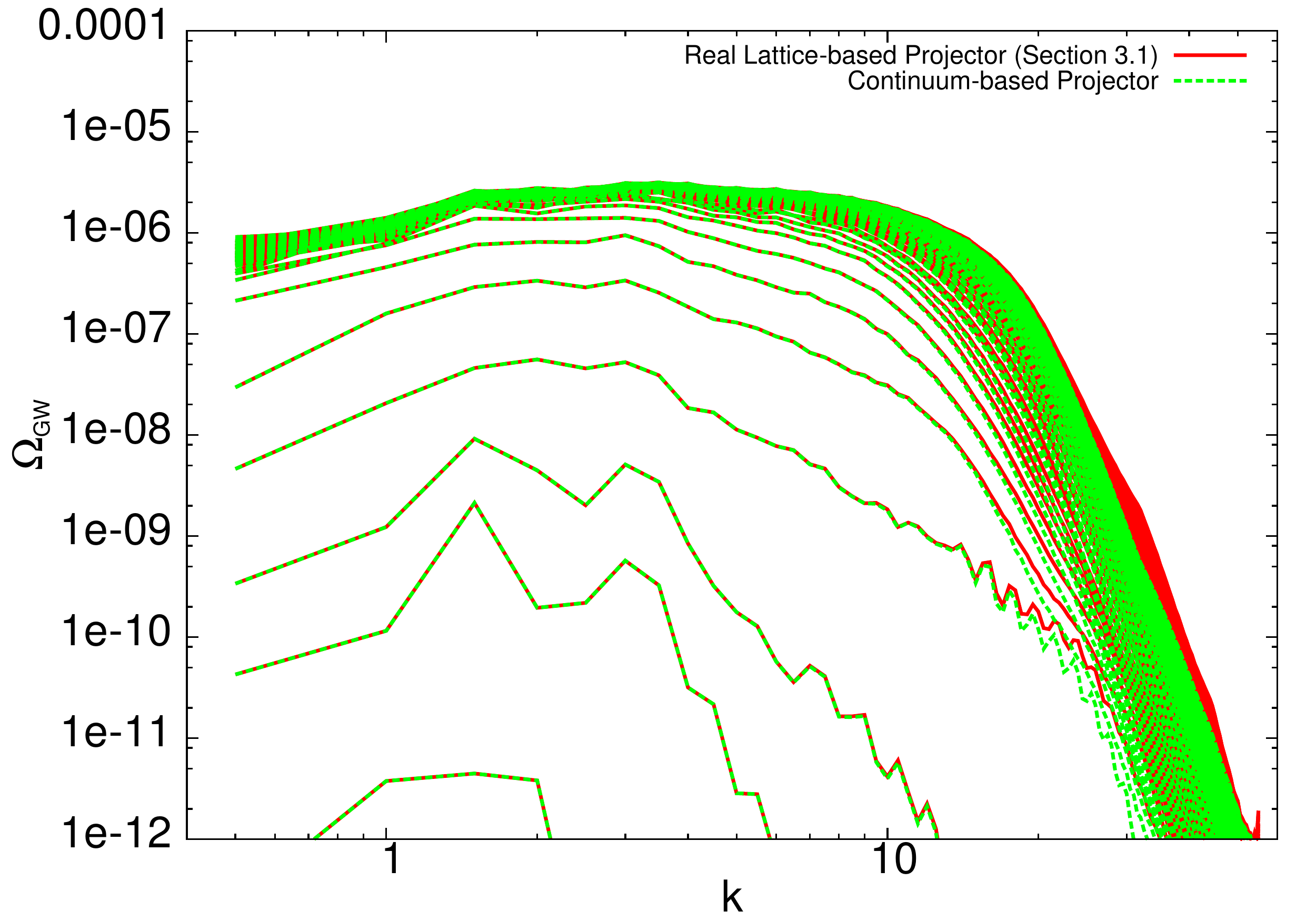}
\epsfig{width=5.5cm,height=7.5cm,angle=-90,file=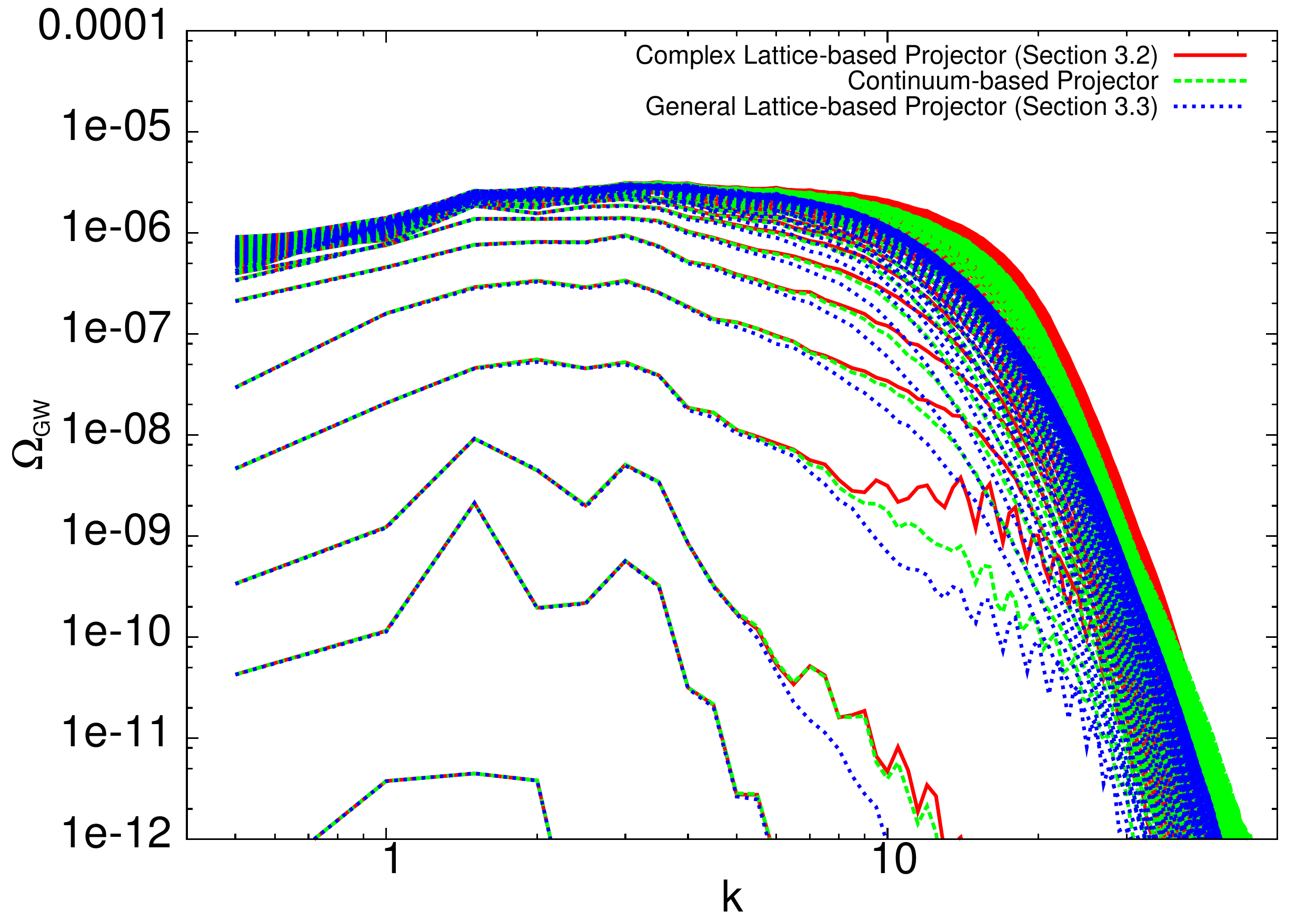}
\caption{The amplitude of the GW spectra obtained with the  different projectors defined in Section~\ref{sec:TTlatticeProjection}. The plots correspond to the evolution of (p)reheating in a Chaotic coupled model $\frac{\lambda}{4}\phi^4 + \frac{g^2}{2}\chi^2\phi^2$, with $\lambda/g^2 = 120$. As in figure~\ref{fig:GW_differenceChaoPure}, we can appreciate that the IR parts of the spectra are identical during the whole evolution, both in the left and right panels. The discrepancies in the UV region are only a factor $\mathcal{O}(1)$ when comparing the output obtained with the continuum-based projector versus the one with the lattice-based real projector of Eq.~(\ref{eq:TTprojectorNeutralDerivatives}), see left panel. When comparing the GW spectra obtained with the continuum-based versus the complex lattice based projectors of eqs.~(\ref{eq:TTprojectorGeneralDerivatives}), (\ref{equ:LambdaFinal}), we see that the difference is more pronounced, a factor $\mathcal{O}(10)$, around the scale at which the UV tail begins to fall exponentially. Nevertheless, the difference in the UV between the two outputs for the different complex projectors, are still less than a factor 3.}
\label{fig:GW_differenceChaoCoupled}
\end{figure}

\begin{figure}
\epsfig{width=5.5cm,height=7.5cm,angle=-90,file=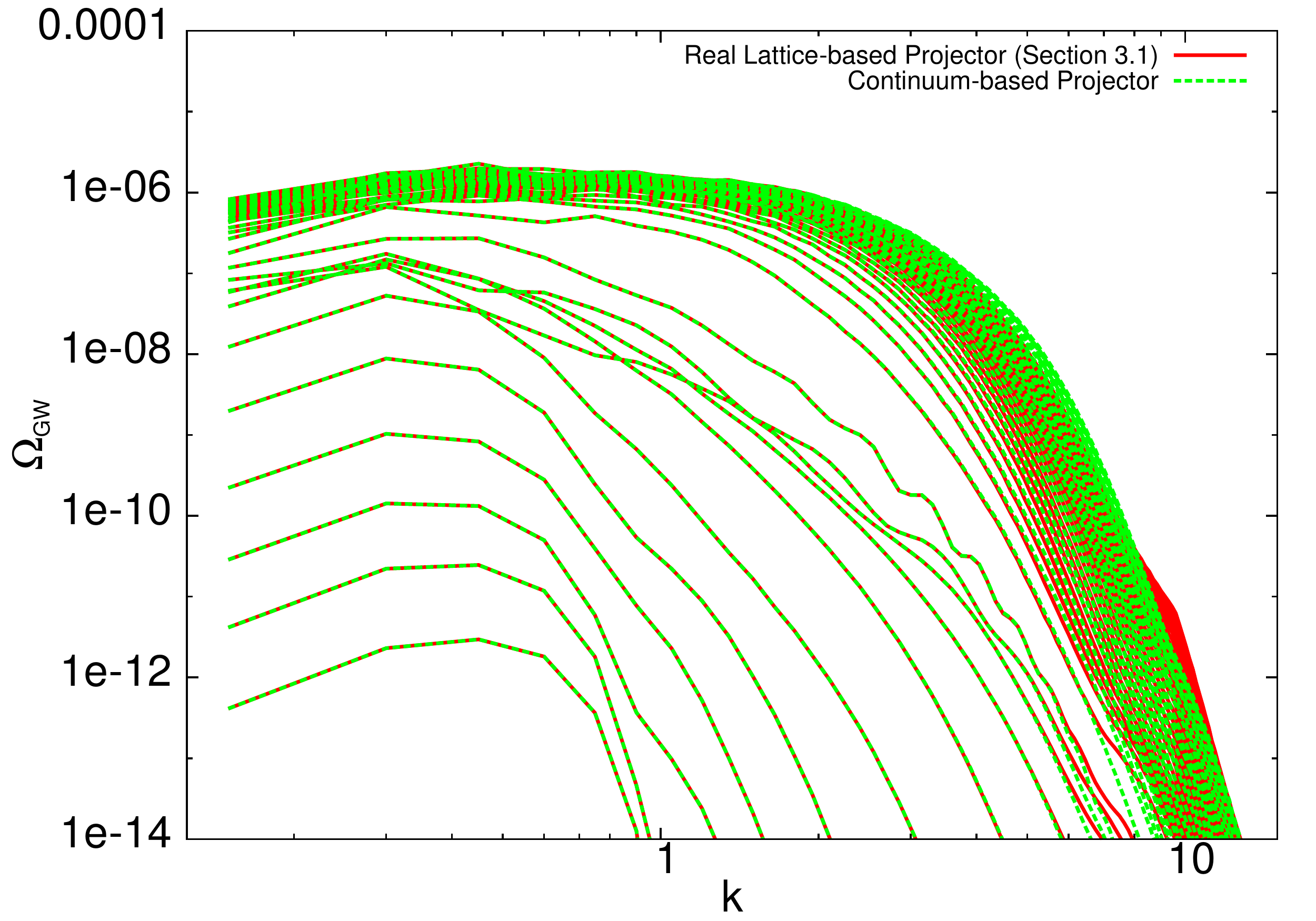}
\epsfig{width=5.5cm,height=7.5cm,angle=-90,file=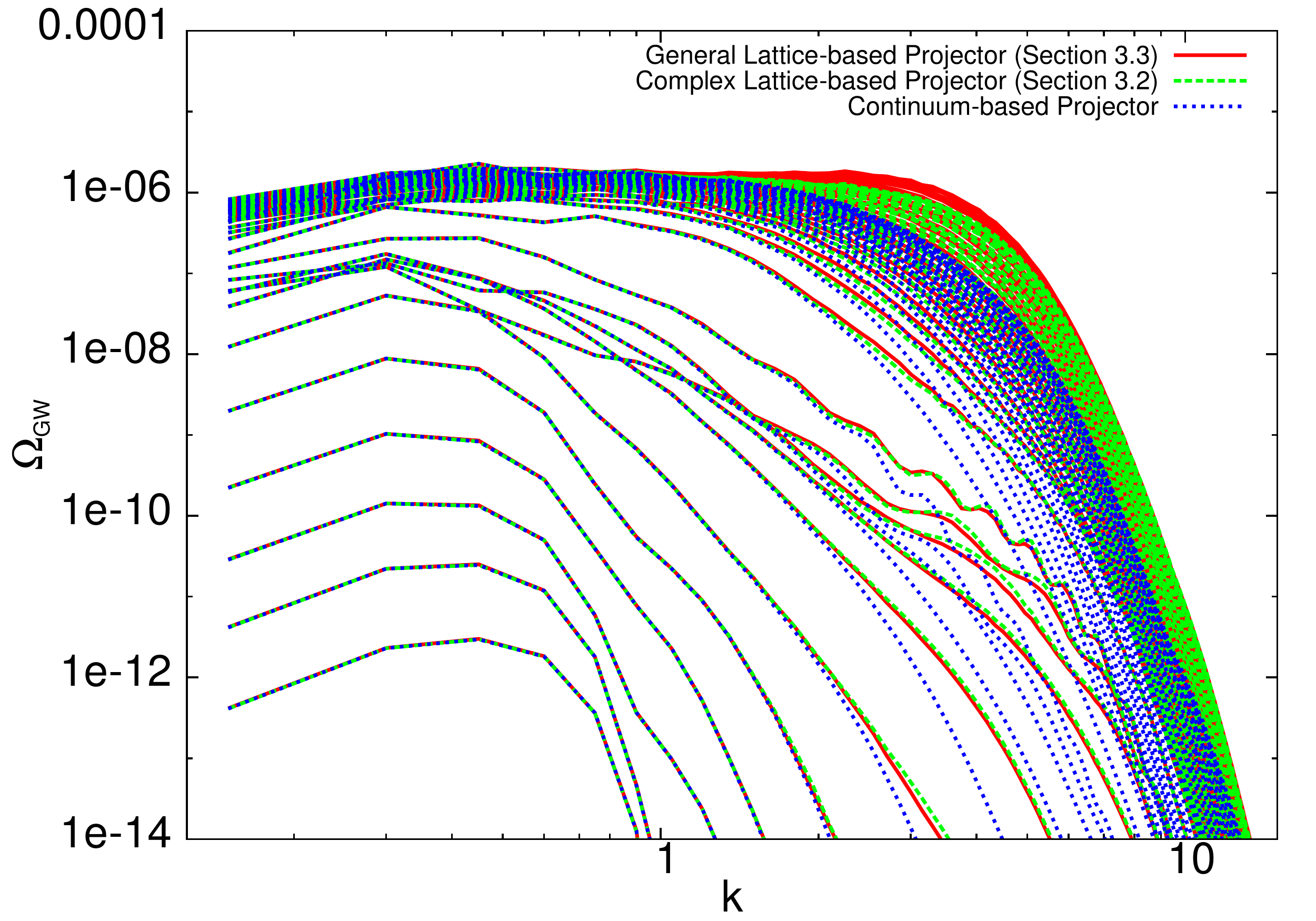}
\caption{The amplitude of the GW spectra obtained, for a Hybrid model, with the different projectors defined in Section~\ref{sec:TTlatticeProjection}. The plots correspond to the evolution after a period of Hybrid inflation described by the model $\lambda(\Phi^2-v^2)^2 + g^2\Phi^2\chi^2$, with $v = 10^{16}$ GeV and $\lambda = 2g^2$. As in previous Figs.~\ref{fig:GW_differenceChaoPure} and~\ref{fig:GW_differenceChaoCoupled}, we can appreciate that the IR parts of the spectra are again identical during the whole evolution, both in the left and right panels. When comparing the output obtained with the continuum-based versus lattice-based real projector of Eq.~(\ref{eq:TTprojectorNeutralDerivatives}), see left panel, the discrepancies in the UV region are only a factor of order one, so again unimportant. However, when comparing the GW spectra obtained with the continuum-based versus the complex lattice-based projectors~(\ref{eq:TTprojectorGeneralDerivatives}), (\ref{equ:LambdaFinal}), the difference in the UV is more noticeable. The discrepancy is a factor $\mathcal{O}(10)$ around the scale at which the UV tail begins to fall exponentially. Nevertheless, the difference in the UV between the two outputs for the different complex projectors, are still less a factor three, similar to the Chaotic coupled scenario.}
\label{fig:GW_differenceHybrid}
\end{figure}

Next we will discuss how the new lattice-based projectors modify the GW spectra as compared to the spectra obtained with the old continuum-based projector. As we show explicitly in Figs.~\ref{fig:GW_differenceChaoPure}, ~\ref{fig:GW_differenceChaoCoupled} and~\ref{fig:GW_differenceHybrid}, the spectra of GW in different models is only modified in the large-momenta region, i.e.~in the ultraviolet (UV) tail. The infrared (IR) features at low momenta, including the shape and amplitude of the spectra, and the position of the peak, are not modified by such UV distortion. The UV region corresponds precisely to those modes for which the GW spectral amplitude should be exponentially suppressed, if the GW spectrum is to be trusted. This is because only the IR modes are excited initially via exponential instabilities during (p)reheating~\cite{preheating,tachyonic,symbreak,symbreak2}, whereas the UV tail of the spectra simply grows by scattering and turbulence~\cite{MichaTkachev}, see for instance Ref.~\cite{EL,GBF,DBFKU} for details. A similar behavior occurs also in the context of gauge fields~\cite{chern,magnetic,CEWB,CLRT,TS,DFGB}.

That the overall shape and final amplitude\footnote{The GW production becomes inefficient in all these models of (p)reheating when the fields enter into the turbulent regime, see~\cite{DBFKU,GBF,EL} for details, so the spectrum amplitude stops growing and saturates to a constant and final shape.} of the GW spectra does not change much when using the new lattice-based projectors, might seem at first sight surprising, given the fact that the degree of transversality changes several orders of magnitude when replacing the continuum-based projector by the lattice-based ones. However, the lattice-momentum $k_{{\rm eff},i}(\tilde\bn)$ from which the lattice-based projectors are made, only differ significantly from the momentum used to build up the continuum-based projector, $k_i = \tilde n_i k_{\rm IR}$, for the highest $\tilde n_i$'s. For instance, ${\rm Re}\lbrace k_{{\rm eff},i}^\pm(\tilde\bn)\rbrace$ = $k_{{\rm eff},i}^0(\tilde\bn)$ = $\sin(2\pi\tilde n_i/N)/dx \approx k_{\rm IR}\tilde n_i + \mathcal{O}(2\pi\tilde n_i/N)^3$ as long as $\tilde n_i/N < 1/2\pi$. Thus, as long as $\tilde n_i$ is not close to the UV boundary of the Fourier-lattice $\tilde n_i = \pm N/2$, and since the $P_{ij}$ operators from which $\Lambda_{ij,lm}$'s are made are quadratic in $k_{{\rm eff},i}$, the difference between the continuum- and the lattice-based projectors can only be proportional to the difference $|k_{{\rm eff},i}(\tilde\bn)|^2-|k_{\rm IR}\tilde\bn_i|^2$, i.e.
\begin{equation}
|\Lambda_{ij,lm}^{\rm cont}(\tilde\bn)-\Lambda_{ij,lm}^{\rm latt}(\tilde\bn)| \sim \mathcal{O}(|\bk_{\rm eff}(\tilde\bn)|^2-|k_{\rm IR}\tilde\bn|^2) \sim \mathcal{O}(2\pi|\tilde\bn|/N)^3  \,.
\end{equation} 

From this point of view, the GW spectra obtained with the lattice-based projectors are not expected to differ much from the spectra obtained with the continuum-based projector. In particular, in the IR region, say $|\tilde\bn| < N/4$, they should be pretty much coincident, the better the smaller $|\tilde\bn|$. Of course, this IR reasoning is still not enough to conclude that the GW spectra with continuum- and lattice-based projectors will not be very different. As small as it might be such difference, if spurious non-TT modes were incorrectly filtered in the continuum-based projector, the difference in amplitude of the two spectra could be enhanced during the dynamical evolution of the fields responsible for the GW production. That is why implementing in a lattice code the new lattice-based projectors is fundamental in order to check whether it makes a difference or not. In Figs.~\ref{fig:GW_differenceChaoPure}, ~\ref{fig:GW_differenceChaoCoupled} and~\ref{fig:GW_differenceHybrid} we quantify this aspect, showing the outcome of numerical simulations in which the TT {\it d.o.f.} are filtered out with the different projectors discussed in Section~\ref{sec:TTlatticeProjection}.

These figures show very nicely the IR aspect just mentioned. Independently of the projector, they all tend asymptotically to the same shape in the IR region, the GW spectra coincide in shape and amplitude at every step of the evolution, independently of the model analyzed. However, the differences in the UV region can be more noticeable, depending on the model. But this also depends on the TT-projector used. For instance, in the left panels of Figures~\ref{fig:GW_differenceChaoPure}, ~\ref{fig:GW_differenceChaoCoupled} and~\ref{fig:GW_differenceHybrid}, we compare the GW amplitude as obtained with the continuum-based projector and with the real lattice-based one defined in Eq.~(\ref{eq:TTprojectorNeutralDerivatives}). The Discrepancies in all models considered are just a factor~$\mathcal{O}(1)$ in the final amplitudes reached, as seen in the left panels of all the Figures. Thus, we can conclude that the difference in the UV tails are totally irrelevant in this case. 

Let us look now at the right hand side panels of the same Figures, where we compare the GW spectral amplitude between the spectra obtained with the continuum-based projector and with the complex lattice-based projectors defined in eqs.~(\ref{eq:TTprojectorGeneralDerivatives}),~(\ref{equ:LambdaFinal}). There we see a more noticeable difference in the UV regions. In particular, we observe that in large range of momenta, the amplitude of the spectra obtained with the lattice-based projectors can be of the order $\mathcal{O}(10)$ bigger than the amplitude for the continuum-based projector. 

In the model ${\lambda\over4}\phi^4$ with no other couplings to secondary fields\footnote{In this model, the GW spectra retain the characteristic peaks of the scalar field power spectra, see Ref.~\cite{GBF} for more details.}, it is worth noting that such discrepancy only appears at a region in $k$-space where the amplitude of the GW spectrum is already suppressed a factor $\mathcal{O}(10^{-5})$ compared to the maximum amplitude. Therefore, in that respect, it is still a marginal discrepancy. 

In a chaotic model $\frac{\lambda}{4}\phi^4 + \frac{g^2}{2}\chi^2\phi^2$, with $\phi$ the inflaton and $\chi$ just another field, the difference in the UV region is however more visible, see right panel of Figure~\ref{fig:GW_differenceChaoCoupled}. The difference is appreciable at scales in which the GW spectra begin to fall off exponentially, but it is not yet suppressed, as compared to the maximum. Such a difference in the amplitude of the GW spectra is appreciable by eye, however it only represents an overall shift of a factor $\mathcal{O}(1)$ in the location of the scale where the UV tail begins to fall. A similar situation arises in the case of a Hybrid model $\lambda(\Phi^2-v^2)^2 + g^2\Phi^2\chi^2$, with $\Phi$ a waterfall field coupled to the inflaton $\chi$. In this scenario, see right panel of Figure~\ref{fig:GW_differenceHybrid}, we find again a discrepancy between the lattice-based and the continuum-based projectors at scales where the spectra is about to fall off exponentially. Such discrepancy represents nevertheless again, only a shift of a factor $\mathcal{O}(1)$ of the scale where the UV tail of the GW spectrum begins to fall.

In any case, both in the hybrid and coupled chaotic models, the difference in the GW spectral amplitude between the output obtained with the two complex lattice-based projectors~(\ref{eq:TTprojectorGeneralDerivatives}) and~(\ref{equ:LambdaFinal}), amounts only to a factor $\mathcal{O}(1)$. So the discrepancies in amplitude among spectra obtained with the two complex projectors are again marginal.

We can conclude that in all the different models considered, the final amplitude, the spectral shape of the IR region and the overall shape of the spectra, are not significantly modified. There is no leak of scalar modes into the amplitude of the GW spectra even if this is obtained with the continuum-based TT-projection, as done repeatedly in the literature. The GW spectra show some difference in the UV region when the spectra are extracted with the complex lattice-based projectors~(\ref{eq:TTprojectorGeneralDerivatives}) and~(\ref{equ:LambdaFinal}). However, those projectors present certain caveats, as discussed in Section~\ref{sec:TTlatticeProjection}, precisely in the the UV region. When comparing the GW spectra obtained with the lattice-based real projector~(\ref{eq:TTprojectorNeutralDerivatives}) and the Continuum-based one, the spectral difference in the UV region is indeed really marginal. In this latter case, the differences in amplitude in all models considered are just a factor $\mathcal{O}(1)$, and only show up in the UV region, where the large-momenta tail of the spectra is already exponentially suppressed.

\section{Conclusions and discussion}
\label{sec:discussion}

In this paper we try to respond to the criticisms made in Ref.~\cite{Huang2011} with respect to the validity of the projectors used in numerical simulations of gravitational wave production at (p)reheating. It was pointed out that the usual procedure employed in order to obtain the Transverse-Traceless part of metric perturbations in lattice simulations was inconsistent with the fact that those fields live in the lattice and not in the continuum. It was claimed that this could lead to a larger amplitude and the wrong shape for the gravitational wave spectra obtained in numerical simulations of (p)reheating, due to the leakage of scalar modes into the tensor (GW) modes. In order to address this issue, we have developed a consistent prescription in the lattice for extracting the TT part of the metric perturbations. We have defined a general complex TT projector based on lattice momenta, as well as a real projector in the lattice. All these projectors satisfy the required symmetry properties associated with gravitational wave amplitudes.

We then run specific numerical simulations of GW production at (p)reheating with the implementation of the various projectors, and demonstrate explicitly that the GW spectra obtained with the old continuum-based TT projection only differ marginally in shape with respect to the new lattice-based projectors. Therefore, we have been able to answer the criticisms of Ref.~\cite{Huang2011} by showing explicitly that the numerical results obtained with the lattice-based projectors do not change appreciably with respect to those with the continuum-based projector. We thus confirm that all previous results in the literature, concerning the spectra of GW coming from lattice simulations of (p)reheating, should be trusted to the extent to which such simulations are trusted (i.e. within lattice artifacts' effects and time evolution). Introducing different lattice momenta in the TT-projector, it only gives rise to differences in the (exponentially suppressed) spectral amplitudes in the UV, or at most to small shifts (or order unity) in the scale where the spectra begin to fall exponentially. The overall shape, frequency and total amplitude do not change.

\begin{table}
\begin{centering}\begin{tabular}{|c|c|c|c|c|}
\hline 
model / projector & continuum & lattice real & latt. complex & latt. general 
\\[+1mm]
\hline \hline
Chaotic Mixed   & $7.125\times10^{-6}$ &  $7.15958\times10^{-6}$ &  $7.42083\times10^{-6}$ & $6.46379\times10^{-6}$\tabularnewline
\hline 
Chaotic Pure   &  $1.8772\times10^{-6} $ &  $1.87652\times10^{-6}$ &  $1.88079\times10^{-6}$ &  $1.86935\times10^{-6}$ \tabularnewline
\hline 
Hybrid         &  $4.6009\times10^{-6}$ &  $4.58099\times10^{-6}$ &  $5.15501\times10^{-6}$ &  $5.70762\times10^{-6}$ \tabularnewline
\hline 
\end{tabular}\par\end{centering}
\caption{The total energy density in gravitational waves in units of the total energy density for the different models and computed using the various projectors. Note that they differ by less than a few percent, except in the hybrid case that can reach 20\% between the continuum and the complex lattice projectors.
\label{tab:rhoGW}}
\end{table}

Finally, in order to quantify by a single number the deviations induced by the choice of projector, to assess the validity of the lattice projector approximation, we have computed the fraction of the total energy density in GW to the total energy density available during (p)reheating. We have included those ratios in Table~\ref{tab:rhoGW}. Assuming that all the energy at the end of inflation went into radiation at reheating, then this fraction represents the fraction of GW to radiation today, since they both redshift equally during the subsequent evolution of the universe. Therefore the quantity that appears in the table must be multiplied by $\Omega_{\rm rad}h^2 = 3.2\times10^{-5}$ in order to give the observable quantity today, $\Omega_{\rm GW}h^2$, which could eventually be measured in gravitational wave observatories. Note that the entries in Tab.~\ref{tab:rhoGW} differ by less than a few percent, except in the hybrid case that can reach 20\% between the continuum and the complex lattice projectors. These numbers reinforce again the idea that the requirement of using the lattice-based projectors does not invalidate the previous results found in the literature on GW production in lattice simulations.

\acknowledgments

DGF would like to express his gratitude to the Theoretical Physics Group at Imperial College London and to the Instituto de F\'isica Te\'orica in Madrid, for the hospitality received during spring/summer 2011, when this project was initiated. This work was supported at Helsinki by the Academy of Finland grant Ref.~131454. AR was supported by the STFC grant ST/G000743/1, and DGF and AR by the Royal Society International Joint Project JP100273. We also acknowledge financial support from the Madrid Regional Government (CAM) under the program HEPHACOS S2009/ESP-1473-02, and MICINN under grant  AYA2009-13936-C06-06. DGF and JGB participate in the Consolider-Ingenio 2010 PAU (CSD2007-00060), as well as in the European Union Marie Curie Network ``UniverseNet" under contract MRTN-CT-2006-035863.

\appendix

\end{document}